\documentclass[letterpaper,english,reprint, aps]{revtex4-1}
\usepackage[T1]{fontenc}
\usepackage[utf8]{inputenc}
\usepackage{color}
\usepackage{amsmath}
\usepackage{amssymb}
\usepackage{graphicx}

\makeatletter

\usepackage[colorlinks = true,
            linkcolor =red,
            urlcolor  =magenta,
            citecolor = blue,
            anchorcolor = blue]{hyperref}
\begingroup

\footnotetext{These authors contributed equally to this work.}
\endgroup

\makeatother

\usepackage{babel}
\begin{document}
\title{Enhancement of non-Gaussianity and nonclassicality of pair coherent
states with postselected von Neumann measurement }
\author{Yi-Fang Ren$^{1}$\textsuperscript{\textsection}, Janarbek Yuanbek$^{1}$\textsuperscript{\textsection}, }
\author{Yusuf Turek$^{1}$}
\email{yusuftu1984@hotmail.com}

\affiliation{$^{1}$School of Physics, Liaoning University, Shenyang, Liaoning
110036, China}
\date{\today}
\begin{abstract}
We investigate the effects of postselected von Neumann measurements
on the nonclassical properties of pair coherent states (PCS). We calculated
key quantum characteristics, such as squeezing, photon statistics,
entanglement between the two PCS modes, and teleportation fidelity.
Our results demonstrate that postselected von Neumann measurements
enhance both the non-Gaussianity and nonclassicality of PCS. These
findings are validated by analyzing the scaled joint Wigner function
across various system parameters. We also notice that in weak measurement
regimes, the teleportation fidelity via a PCS-based quantum channel
under postselected von Neumann measurements could keep successful
teleportation for anomalous weak values of the system obervable. The
theoretical optimization scheme offers an alternative approach for
improving PCS-based quantum information efficiency and facilitates
practical implementations in quantum technologies. 
\end{abstract}
\maketitle

\section{Introduction\label{sec:1}}

The nonclassicality of quantum states plays a crucial role in quantum
computation and information processing. Recent advancements in technology
allow the manipulation of photons, making continuous-variable states
of bosonic modes an excellent platform for quantum information processing.
Among the variety of continuous-variable states \citep{2020quantum,PRXQuantum.2.030204,PhysRevA.109.040101},
two-mode Gaussian and non-Gaussian states such as entangled cat states
\citep{doi:10.1126/science.aaf2941}, N00N states \citep{doi:10.1126/science.1188172,PhysRevLett.106.060401},
squeezed states \citep{Daoming2015QuantumPO,2021}, and pair coherent
states (PCS) \citep{PhysRevLett.57.827} have irreplaceable roles
in numerous applications \citep{PhysRevA.50.2870,doi:10.1142/S0219749907002475,RN171,Seshadreesan_2015,PhysRevLett.117.110801,Wang2017,PhysRevLett.123.231108,RN168,PhysRevResearch.6.043113}.
In two-mode nonclassical states, enhancing nonclassicality and entanglement
properties can be achieved through photon addition and subtraction
operations \citep{HONG1999265,Lu,PhysRevA.73.042310,PhysRevA.75.032104,RN4,RN1,doi:10.1142/S0217979216500326,YUAN20181034,RN2,Duc_2020,RN5,RN170,Chuong2023}.
However, operating with non-Gaussian states introduces additional
computational complexity compared to Gaussian states. Beyond photon
addition and subtraction (or their superpositions) \citep{2007PV},
quantum catalysis provides another feasible method for enhancing the
nonclassicality of quantum states \citep{RN5}. Quantum catalysis
enhances quantum properties without directly altering the photon count.
This enhancement arises from the inherently quantum mechanical nature
of the process (for further details, refer to \citep{PhysRevLett.96.083601},
and references therein). However, quantum catalysis is not universally
applicable to all states, necessitating alternative methods for optimizing
nonclassical states. Specifically, exploring new state-optimization
schemes is essential for improving the efficiency of multimode state-based
applications–such as multimode Gaussian and non-Gaussian states–without
relying on photon addition or subtraction operations. 

The process of optimizing quantum states relates to the measurement
problems. It requires modifying and effectively manipulating measurement
procedures to enhance the properties of a given state using quantum
control techniques \citep{2014Quantum}. Quantum weak measurements
(WMs) \citep{AharonovPhysRevLett.60.1351}, a novel class of generalized
measurement methods, offer feasible and effective pathways for optimizing
quantum states for specific tasks. Unlike the strong measurements,
WMs involve a weak coupling between the measured system and the measuring
device, which prevents the destruction of the system's initial state.
While a single trial of a WM does not yield precise information about
the system, repeated measurements combined with postselection allow
for the statistical extraction of desired observable information.
The extracted weak value is a complex quantity that can lie outside
the eigenvalue spectrum of the observable \citep{Tamir2013,WeakMeasurements}.
This phenomenon, known as weak value amplification (WVA), is useful
for amplifying weak signals in quantum systems \citep{PhysRevLett.118.070802}
(for further insights into WM theory and its applications, readers
may consult the works of Nori \citep{70} and Boyd \citep{71}). The
utility of quantum WM techniques in state preparation \citep{PhysRevA.105.022608}
and optimization processes has been demonstrated for various quantum
states and optimization processes has been demonstrated for various
quantum states \citep{PhysRevA.105.022210,YusufRN148}. However, their
advantages for multimode continuous-variable states of bosonic modes
remain largely unexplored. For instance, while single-mode coherent
states are typical semiclassical states that minimize (and equalize)
the uncertainty product of two incompatible operators, PCS serve as
two-mode analogs of Glauber coherent states. PCS are eigenstates of
the pair-photon annihilation operator and represent a promising candidate
for applying quantum WM-based state optimization methods. 

In this work, we investigate how the properties of PCS change after
applying the theory of postselected measurements to one mode of PCS.
We calculate the squeezing, quantum statistics, and entanglement characteristics
of the measurement output state. Our results show that postselected
von Neumann measurements could enhance the non-Gaussianity and nonclassicality
of PCS, considering large weak values of the measured system observable
and appropriate coupling strength parameters between the measured
system and measuring device (MD). We confirm these findings by examining
the phase-space distribution of our state, characterized by the scaled
joint Wigner function. We also check the teleportation fidelity with
postselected von Neumann measurements to the PCS and the numerical
outcomes indicate that in weak measurement regimes the quantum teleportation
task based on PCS-based quantum channel is still could keep its validity
for larger weak values of measured system observable. 

We organized the paper as follows: In Sec. \ref{sec:MODEL-SETUP},
we present our theoretical model. Section \ref{sec:The-effects squeezing}
explores the effects of postselection on the squeezing properties
of PCS, including quadrature and sum squeezing. In Sec. \ref{sec:The-effects-quantum statics},
we investigate the quantum statistics of postselected PCS under von
Neumann measurements, analyzing the relationship between the $a$
mode and $b$ mode using the second-order cross-correlation (SOCC)
and second-order correlation functions. Section \ref{sec:The-effects entanglement}
examines entanglement through measures such as the Hillery-Zubairy
(HZ) correlation and Einstein–Podolsky–Rosen (EPR) correlation. In
Sec. \ref{sec:6}, we investigate the scaled joint Wigner function
of our measurement output state and take comparison with the initial
PCS case. Section \ref{sec:Fidelity} we present the state distance
between initial and final MD states, and give the details of the teleportation
fidelity of a coherent state transmitted through a PCS-based quantum
channel under postselected von Neumann measurements. Finally, Sects.
\ref{sec:Discussion} and \ref{sec:conclusion} present a discussion
and conclusion, respectively. Throughout this paper, we take $\hbar=1$.

\section{\label{sec:MODEL-SETUP}MODEL SETUP}

The coherent state is a typical semi-classical state that minimizes
(and equalizes) the uncertainty product of two incompatible operators.
However, the Gaussian profile characterizing the classicality of coherent
states transforms into a non-Gaussian profile upon the addition of
a single photon. Beyond the standard coherent state and the single-photon-added
coherent state, there also exist two-mode entangled coherent states,
commonly referred to as PCS \citep{Agarwal:88}.

Analogous to the usual coherent state, defined as $\vert\alpha\rangle=e^{-\vert\alpha\vert^{2}/2}\sum_{n}\alpha^{n}/\sqrt{n!}\vert n\rangle$,
the PCS are a superposition of twin-Fock state bases $\vert n+\delta\rangle_{a}\vert n\rangle_{b}$
of two harmonic oscillators ($a$ and $b$): 

\begin{equation}
|\gamma,\delta\rangle=\mathcal{N}_{\delta}\sum_{n=0}^{\infty}\frac{\gamma^{n+\delta/2}}{\sqrt{n!(n+\delta)!}}\left|n+\delta\right\rangle _{a}\left|n\right\rangle _{b}.\label{eq:fock}
\end{equation}
Here, $\gamma$ is a complex number representing the amplitude and
phase of the state, $\delta$ is an integer denoting the photon number
difference (PND) between the two modes; and $\mathcal{N}_{\delta}$
is a normalization coefficient defined by: 

\begin{align}
\mathcal{N}_{\delta} & =\left(\sum_{n=0}^{\infty}\frac{|\gamma|^{2n+\delta}}{n!(n+\delta)!}\right)^{-\frac{1}{2}}=\frac{1}{\sqrt{I_{\delta}(2|\gamma|)}}.\label{eq:2-1}
\end{align}
Here, $I_{\delta}(2|\gamma|)$ denotes the modified Bessel function
of the first kind, expressed as follows: 
\begin{align}
I_{\delta}(x) & =\sum_{n=0}^{\infty}\frac{1}{n!(n+\delta)!}\Big(\frac{x}{2}\Big)^{2n+\delta}.\label{eq:3}
\end{align}
This state is an eigenstate of the pair-photon annihilation operator
$ab$ and the PND operator $a^{\dagger}a-b^{\dagger}b$: 
\begin{align}
ab\vert\gamma,\delta\rangle & =\gamma\vert\gamma,\delta\rangle,\\
\left(a^{\dagger}a-b^{\dagger}b\right)\vert\gamma,\delta\rangle & =\delta\vert\gamma,\delta\rangle.
\end{align}
Here, $a$ ($a^{\dagger}$) and $b$ ($b^{\dagger}$) are the annihilation
(creation) operators associated with the two modes ($a$ mode and
$b$ mode) of the PCS, respectively. The various generation methods
have been proposed theoretically in different physical systems \citep{PhysRevLett.57.827,PhysRevA.54.4315,RN165,OBADA200619,DONG20085677,PhysRevA.84.023810}.
The generation and characterization of PCS within spatially two confined
superconducting cavities have been explored in recent experimental
work \citep{ExperimentalPCSPRXQuantum.4.020319}. As indicated in
previous theoretical \citep{RN165} and experimental \citep{RN172}
works, it also may possible to generate and manipulate of PCS in trapped
ions systems. In order to facilitate the PCS in long-distance quantum
information processing and communication tasks in most recent work
\citep{Truong_2024} the experimentally feasible theoretical proposal
to the generation of PCS in open space with high fidelity also introduced
in detail. PCS are a type of two-mode non-Gaussian and nonclassical
states \citep{Lu}, and its sub-Poissonian statistics, correlations
in number fluctuations, squeezing, and violations of Cauchy–Schwarz
inequalities (CSI) have been studied extensively \citep{Agarwal:88,PhysRevA.41.1569,Obada_2020,Duc_2020,RN170}.
As non-Gaussian entangled states, PCS offers advantages in performing
quantum tasks such as quantum communication \citep{PhysRevA.50.2870,Quantumteleportation,Albert_2019},
interferometric lithography \citep{PhysRevA.88.043841}, quantum metrology
\citep{PhysRevLett.96.010401}, quantum dialogue \citep{NGUYEN20046}
and quantum key distribution \citep{RN171,Wang2017}.

\begin{figure}
\includegraphics[width=8cm]{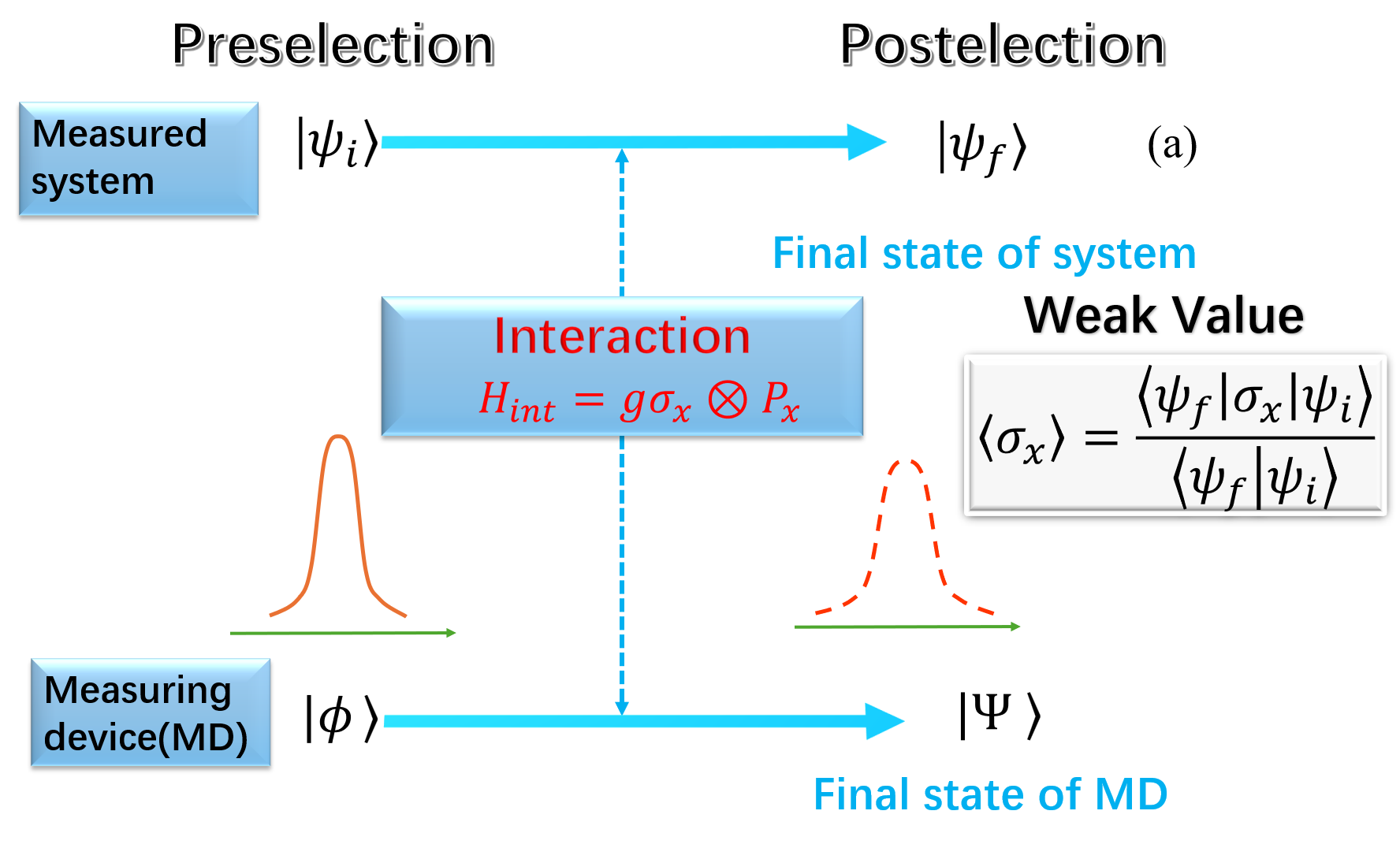}

\includegraphics[width=8cm]{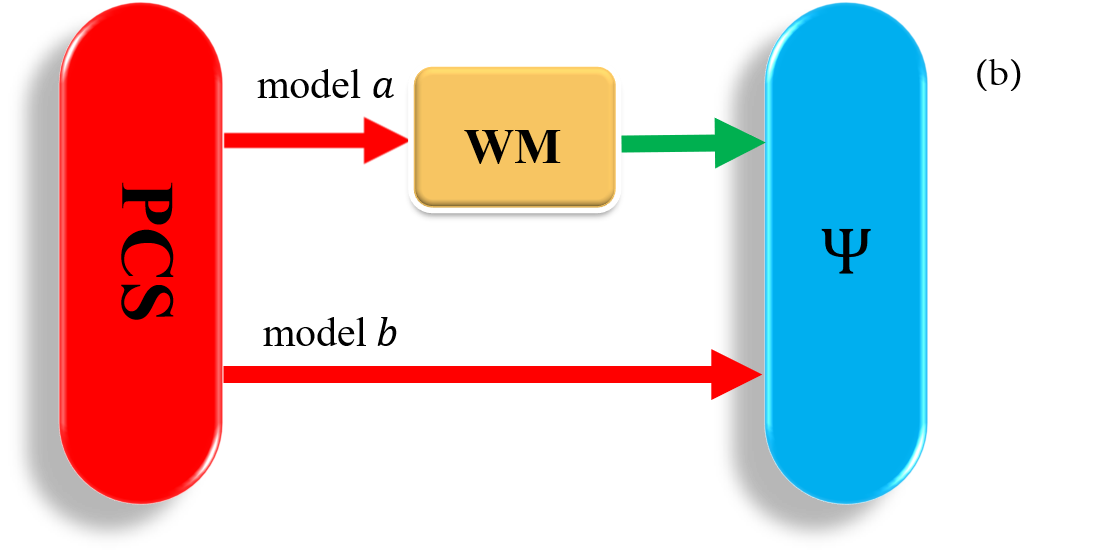}

\caption{\label{fig:1Theoretical-Schematic}(a) Schematic diagram of weak measurement
(WM) theory. The standard process of WM involves four steps: 1. \textit{Preparation:
}The initial state of the measured system is prepared as $\vert\psi_{i}\rangle$,
while the MD is initialized in the state $\vert\phi\rangle$. 2. \textit{Weak
Interaction:} A weak interaction occurs between the measured system
and the MD, during which the composite system evolves according to
the Hamiltonian $H_{int}=g\,\sigma_{x}\otimes P_{x}$. Here, the coupling
constant $g$ characterizes the bilinear coupling. 3. \textit{Postselection:}
After evolution, the entire system is projected onto the postselected
state $\vert\psi_{f}\rangle$. This step extracts the desired observable
values of the system by selecting a specific subensemble of samples
before the final measurement. 4. \textit{Readout:} The measurement
result determines the shifts in the MD. In the postselected weak measurement
process, the observable value is a function of the weak value. The
real ($Re$) and imaginary ($Im$) parts of the weak value are derived
from the shifts in the canonical position and momentum of the MD respectively.
(b) Schematic setup for preparing $\vert\Psi\rangle$ via postselected
von Neumann measurement. Experimentally, the pointer is initially
in the PCS. Using the WM strategy, where the interaction is restricted
to the $a$-mode, we investigate the effects of this measurement on
the properties of the PCS.}
\end{figure}

To study the effects of postselected von Neumann measurement on the
PCS, we consider the PCS as the MD and its polarization as the measured
system. As illustrated in Fig. \ref{fig:1Theoretical-Schematic},
we simplify the setup by restricting the interaction between the MD
and the measured system to the $a$-mode. The interaction Hamiltonian
$H_{int}$ is of the von Neumann type, expressed as: 

\begin{equation}
H_{int}=g\sigma_{x}\otimes P_{x}.\label{eq:interaction Hammiltonian}
\end{equation}
Here, $g$ denotes the interaction strength between the MD and the
measured system. The operator $\sigma_{x}=\vert D\rangle\langle D\vert-\vert A\rangle\langle A\vert$
represents the observable of the measured system, where $\vert D\rangle=\frac{1}{\sqrt{2}}\left(\vert H\rangle+\vert V\rangle\right)$,
and $\vert A\rangle=\frac{1}{\sqrt{2}}\left(\vert H\rangle-\vert V\rangle\right)$
correspond to the diagonal and anti-diagonal polarizations of the
beam, respectively. Here, $\vert H\rangle$ and $\vert V\rangle$
denote the horizontal and vertical polarizations of photons. In the
interaction Hamiltonian, $P_{x}$ represents the canonical momentum
of the MD and can be expressed in terms of annihilation and creation
operators as follows: 
\begin{equation}
P_{x}=\frac{i}{2\sigma}\left(a^{\dagger}-a\right),\label{eq:2}
\end{equation}
where $\sigma=\sqrt{1/2m\omega}$ represents the beam width. We now
assume that the composite system is initially prepared in the following
form: 
\begin{equation}
\vert\Psi_{in}\rangle=\vert\psi_{i}\rangle\otimes\vert\phi\rangle,\label{eq:intial state}
\end{equation}
where $\vert\psi_{i}\rangle=\cos\frac{\alpha}{2}\vert H\rangle+e^{i\vartheta}\sin\frac{\alpha}{2}\vert V\rangle$
and $\vert\phi\rangle=|\gamma,\delta\rangle$. As shown in Fig. \ref{fig:1Theoretical-Schematic},
$\vert\psi_{i}\rangle$ and $\vert\phi\rangle$ represent the states
of the measured system and the MD, respectively. The initial state
is thus the product state between the measured system and the MD.
Here, $\vartheta\in[0,2\pi]$ and $\alpha\in[0,\pi)$. Experimentally,
this type of initial state can be prepared by passing the beam through
a quarter-wave and half-wave plates with appropriate optical axis
angles. The joint state, expressed in Eq. (\ref{eq:intial state}),
evolves over time under the influence of the time evolution operator
$U(t)=\exp\left(-i\int_{0}^{t}H_{int}d\tau\right)$, as follows: 
\begin{eqnarray}
\vert\Psi_{evol}\rangle & = & \exp\left(-i\int_{0}^{t}H_{int}d\tau\right)\vert\Psi_{in}\rangle\nonumber \\
 & = & \frac{1}{2}\left[(\mathbb{I}+\sigma_{x})D\left(\frac{\Gamma}{2}\right)+(\mathbb{I}-\sigma_{x})D^{\dagger}\left(\frac{\Gamma}{2}\right)\right]\label{eq:time evolution}\\
 & \times & \vert\psi_{i}\rangle\otimes\vert\phi\rangle,\nonumber 
\end{eqnarray}
where $\Gamma=gt/\sigma$ represents the ratio between $gt$ and $\sigma$,
$\mathbb{I}$ is the $2\times2$ identity matrix, and $D(\Gamma/2)=e^{\frac{\Gamma}{2}(\hat{a}^{\dagger}-\hat{a})}$
is the displacement operator acting on the $a$ mode of the PCS. It
is important to note that the coupling strength parameter $\Gamma$
characterizes the measurement strength in this study. The coupling
between measured system and MD is referred to as weak when $\Gamma\text{\ensuremath{\ll1}}$
and strong when $\Gamma\gg1$. For this reason, $\Gamma$ can be considered
the transaction parameter in our measurement model. The measurement
process involves postselection. In this study, we assume the postselection
of the system state to be $\vert\psi_{f}\rangle=\vert H\rangle$.
After postselecting the system state, the final state of the MD is
obtained, and its expression is given as 

\begin{align}
\vert\Psi\rangle & =\frac{\lambda}{2}\left[\left(1+\langle\sigma_{x}\rangle_{w}\right)D\left(\frac{\Gamma}{2}\right)+\left(1-\langle\sigma_{x}\rangle_{w}\right)D^{\dagger}\left(\frac{\Gamma}{2}\right)\right]\vert\phi\rangle.\label{eq;postselection}
\end{align}
Here, $\lambda$ is the normalization coefficient given by 
\begin{equation}
\lambda=\frac{\sqrt{2}}{\left[1+|\langle\sigma_{x}\rangle_{w}|^{2}+(1-|\langle\sigma_{x}\rangle_{w}|^{2})P\right]^{\frac{1}{2}}},
\end{equation}
with $P=\mathcal{N}_{\delta}^{2}e^{-\frac{\Gamma^{2}}{2}}\sum\limits_{n=0}^{\infty}\left[\frac{|\gamma|^{2n+\delta}}{n!(n+\delta)!}L_{n+\delta}^{(0)}\left(\Gamma^{2}\right)\right]$,
and $\langle\sigma_{x}\rangle_{w}$ is the weak value of the system
observable $\sigma_{x}$ defined as
\begin{align}
\langle\sigma_{x}\rangle_{w} & =\frac{\langle\psi_{f}\vert\sigma_{x}\vert\psi_{i}\rangle}{\langle\psi_{f}\vert\psi_{i}\rangle}=e^{i\vartheta}\tan\frac{\alpha}{2}.\label{eq:weak value}
\end{align}
For simplify, we assume the weak value to be a real number by setting
$\vartheta=0$. In calculating the normalization coefficient $\lambda$,
the following expressions were used 
\begin{align}
\langle n+d|D(\alpha)|n\rangle & =\sqrt{\frac{n!}{(n+d)!}}e^{-\frac{|\alpha|^{2}}{2}}\alpha^{d}L_{n}^{\left(d\right)}\left(|\alpha|^{2}\right),\\
\langle n|D(\alpha)|n+d\rangle= & \sqrt{\frac{n!}{(n+d)!}}e^{-\frac{|\alpha|^{2}}{2}}(-\alpha^{*})^{d}L_{n}^{\left(d\right)}(|\alpha|^{2}),
\end{align}
where $d$ is natural number, and $L_{n}^{\left(d\right)}\left(|\alpha|^{2}\right)$
represents the associated Laguerre polynomials, given by 
\begin{equation}
L_{n}^{(d)}\left(|\alpha|^{2}\right)=\sum_{m=0}^{n}\left(-1\right)^{m}C_{n+d}^{m+d}\frac{|\alpha|^{2m}}{m!}.
\end{equation}

We have to mention that in the derivation of final state of MD, i.e.,
Eq. (\ref{eq;postselection}), we don't use any approximation and
the coupling strength parameter $\Gamma$ can take arbitrary value.
If we interested in standard weak measurement regime, its enough to
us only consider up to first-order term of the displacement operator
$D\left(\frac{\Gamma}{2}\right)$ since $\Gamma\ll1$ in this case,
i.e., $D\left(\frac{\Gamma}{2}\right)\approx1+$$\frac{\Gamma}{2}(a^{\dagger}-a)$.
However, as studied in previous works \citep{Turek_2015,PhysRevA.92.022109},
the weak value $\langle\sigma_{x}\rangle_{w}$ in our scheme just
a quantity of system observable $\sigma_{x}$ induced by particular
pre- and post-selection processes on measured system, and it's valid
both in Aharonov’s weak measurement and von Neumann’s projective strong
measurement regimes. 

Next, we examine the effects of postselected von Neumann measurement
and the weak values of the measured system’s observable on the inherent
properties of the PCS $\vert\phi\rangle$.

\section{\label{sec:The-effects squeezing}The effects on squeezing }

In this section, we investigate the effects of postselected von Neumann
measurement on the squeezing properties of the PCS by examining both
quadrature squeezing and sum squeezing.

\subsection{Quadrature squeezing}

In continuous-variable quantum optics, quadrature squeezing is a crucial
characteristic of nonclassical radiation fields, playing a vital role
in the implementation of various quantum computation and communication
protocols \citep{2017A}. Here, we analyze how postselected von Neumann
measurement influences the quadrature squeezing of the PCS.

For a single-mode radiation field, squeezing refers to the reduction
of the quadrature variance below the shot noise level, i.e., $\triangle^{2}X_{\epsilon}<\frac{1}{4}$.
The quadrature operator with phase $\epsilon$ is defined as: $X_{\epsilon}=\frac{1}{2}\left(e^{-i\epsilon}a+e^{i\epsilon}a^{\dagger}\right)$,
where its variance is given by: $\triangle^{2}X=\langle X^{2}\rangle-\langle X\rangle^{2}$.
Similarly, for two-mode radiation fields, the quadrature operators
can be defined as \citep{SCHNABEL20171}: 

\begin{equation}
F_{1}=\frac{1}{2^{3/2}}\left[e^{-i\epsilon}(a+b)+e^{i\epsilon}(a^{\dagger}+b^{\dagger})\right],\label{eq:16}
\end{equation}
\begin{equation}
F_{2}=\frac{1}{2^{3/2}i}\left[e^{-i\epsilon}(a+b)+e^{i\epsilon}(a^{\dagger}+b^{\dagger})\right].\label{eq:17}
\end{equation}
They satisfy the commutation relation $\left[F_{1},F_{2}\right]_{~}=\frac{i}{2}$,
and the uncertainty relation for their fluctuations is 
\begin{equation}
\Delta^{2}F_{1}\Delta^{2}F_{2}\geq\frac{1}{16},\label{eq:18}
\end{equation}
where $\Delta F_{i}^{2}=\langle F_{i}^{2}\rangle-\langle F_{i}\rangle^{2}$.
Similar to the single-mode case, two-mode squeezing occurs when one
of the variances $\triangle^{2}F_{i}$ is below the shot noise level,
i.e., $\triangle^{2}F_{i}<0.25$ ($i=1,2$). This condition can be
satisfied if the two modes are uncorrelated, with one or both of $F_{i}$
individually squeezed, or when nonclassical correlations, such as
entanglement between the two modes, are present. We use the squeezing
parameter to characterize the squeezing of the quadrature as follows:
\begin{equation}
Q_{i}=\Delta F_{i}^{2}-\frac{1}{4}.\label{eq:19}
\end{equation}
As we can see, the values of $Q_{i}$ are bounded as $Q_{i}\geq-\frac{1}{4}$,
and the $i$-th component of the quadrature operators of the PCS is
said to be squeezed if $-\frac{1}{4}\le Q_{i}<0$. After some algebra,
we can obtain the specific expressions for the quadrature squeezing
parameter $Q_{i}$ of the final MD state $\vert\Psi\rangle$, expressed
as: 
\begin{align}
Q_{1,\Psi} & =\frac{1}{4}\left[\langle a^{\dagger}a\rangle+\langle b^{\dagger}b\rangle+Re\left[\langle a^{2}\rangle\right]+Re\left[\langle b^{2}\rangle\right]\right]\nonumber \\
 & +\frac{1}{2}\left[Re\left[\langle ab\rangle\right]+Re\left[\langle a^{\dagger}b\rangle\right]\right]\nonumber \\
 & -\frac{1}{2}\left[Re\left[\langle a\rangle\right]+Re\left[\langle b\rangle\right]\right]^{2},\label{eq:20}
\end{align}
 and

\begin{align}
Q_{2,\Psi} & =\frac{1}{4}\left[\langle a^{\dagger}a\rangle+\langle b^{\dagger}b\rangle+Re\left[\langle a^{2}\rangle\right]+Re\left[\langle b^{2}\rangle\right]\right]\nonumber \\
 & -\frac{1}{2}\left[Re\left[\langle ab\rangle\right]-Re\left[\langle a^{\dagger}b\rangle\right]\right]\nonumber \\
 & -\frac{1}{2}\left[Im\left[\langle a\rangle\right]+Im\left[\langle b\rangle\right]\right]^{2}.\label{eq:21}
\end{align}

Here, $\langle\dots\rangle$ represents the expectation value of the
associated operators under the the final MD state $\vert\Psi\rangle$,
while $Re$ and $Im$ denote its real and imaginary parts. Additionally,
we set $\epsilon=0$. The Appendix \ref{sec:A1} provides analytic
expressions for the expectation values of the associated operators.
If we set $\Gamma=0$, the above two expressions reduce to $Q_{1,\phi}$
and $Q_{2,\phi}$, corresponding to the initial MD state $\vert\phi\rangle$
defined in Eq. (\ref{eq:fock}), and expressed as
\begin{align}
Q_{1,\phi} & =\frac{1}{4}\left(\frac{|\gamma|I_{\delta-1}(2|\gamma|)}{I_{\delta}(2|\gamma|)}+\frac{|\gamma|I_{\delta+1}(2|\gamma|)}{I_{\delta}(2|\gamma|)}+2Re[\gamma]\right),\label{eq:22}
\end{align}
 and 
\begin{equation}
Q_{2,\phi}=\frac{1}{4}\left(\frac{|\gamma|I_{\delta-1}(2|\gamma|)}{I_{\delta}(2|\gamma|)}+\frac{|\gamma|I_{\delta+1}(2|\gamma|)}{I_{\delta}(2|\gamma|)}-2Re[\gamma]\right),\label{eq:23}
\end{equation}
Here, $I_{\delta}(2|\gamma|)$ is the modified Bessel function defined
in Eq. (\ref{eq:3}). From $Q_{1,\phi}$, we can deduce that $Q_{1,\phi}\ge0$
for all $\gamma$, indicating that the $F_{1}$ quadrature of initial
PCS $\vert\phi\rangle$ never exhibits squeezing. Furthermore, if
the third term, $2Re[\gamma]$, in $Q_{2,\phi}$ exceeds the sum of
the first and second terms, its value becomes negative, leading to
the possibility of quadrature squeezing in $F_{2}$. 

\begin{figure}[h]
\includegraphics[width=8cm]{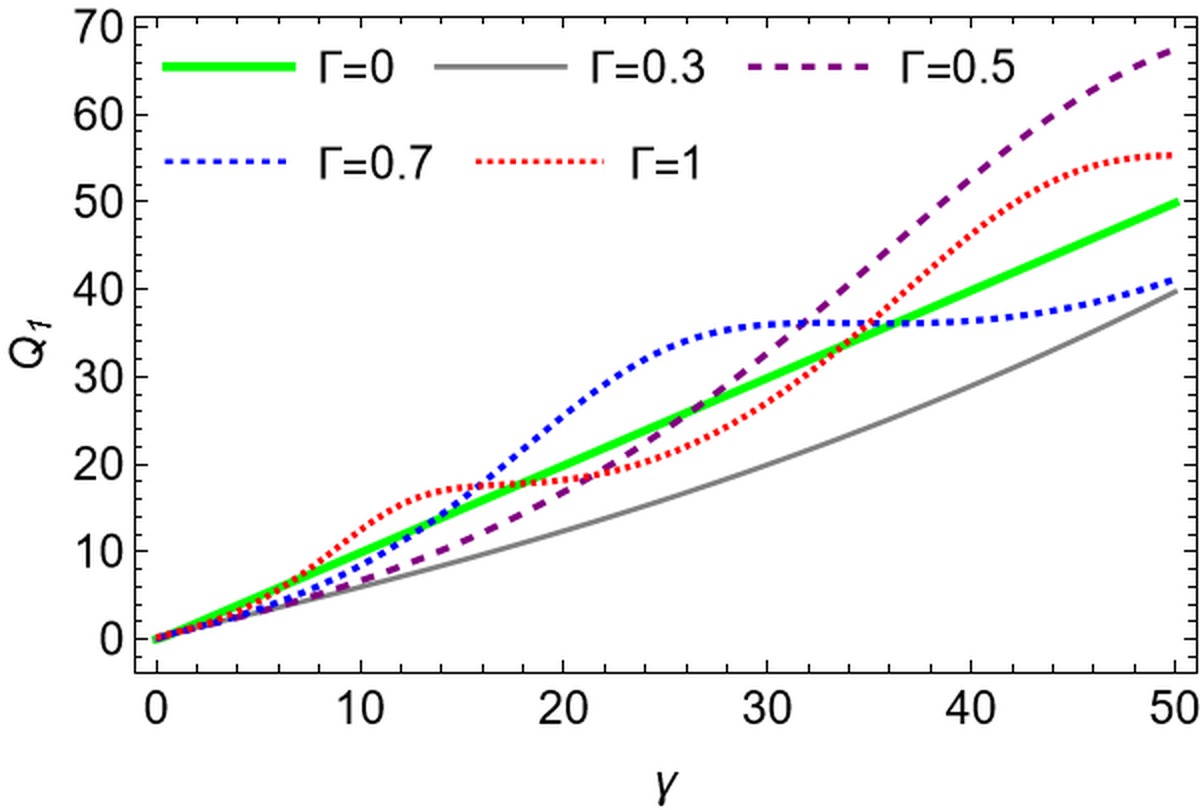}

\caption{\label{fig:2}Quadrature squeezing $Q_{1,\Psi}$ as a function of
the \textcolor{blue}{state} parameter $\gamma$. The thick green line
corresponds to the \textcolor{blue}{coupling strength} parameter $\Gamma=0$,
representing the case where the state is the initial PCS $\vert\phi\rangle$.
The solid gray line represents $\Gamma=0.3$, while the purple, blue,
and red dashed lines correspond to $\Gamma=0.5$, $\Gamma=0.7$, and
$\Gamma=1$, respectively. For these calculations, we take $\delta=0$,
$\alpha=\frac{8\pi}{9}$ and $\vartheta=0$ . }
\end{figure}

\begin{figure}[h]
\includegraphics[width=8cm]{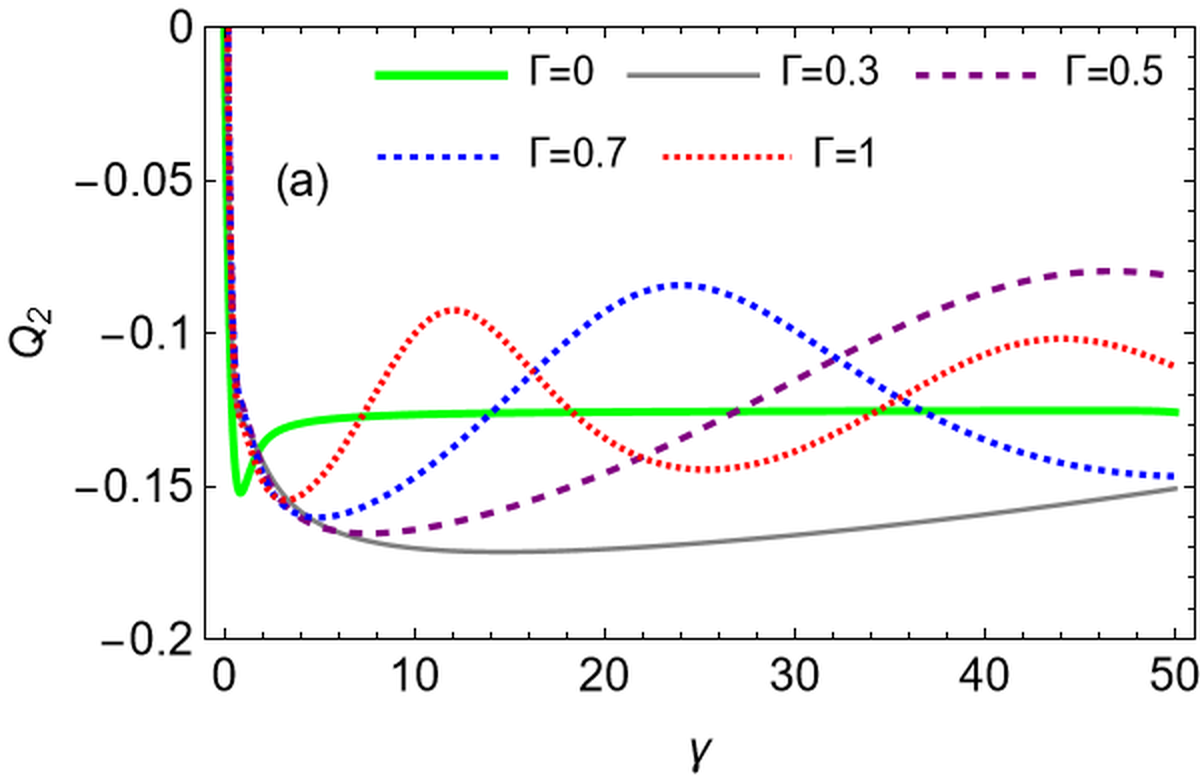}

\includegraphics[width=8cm]{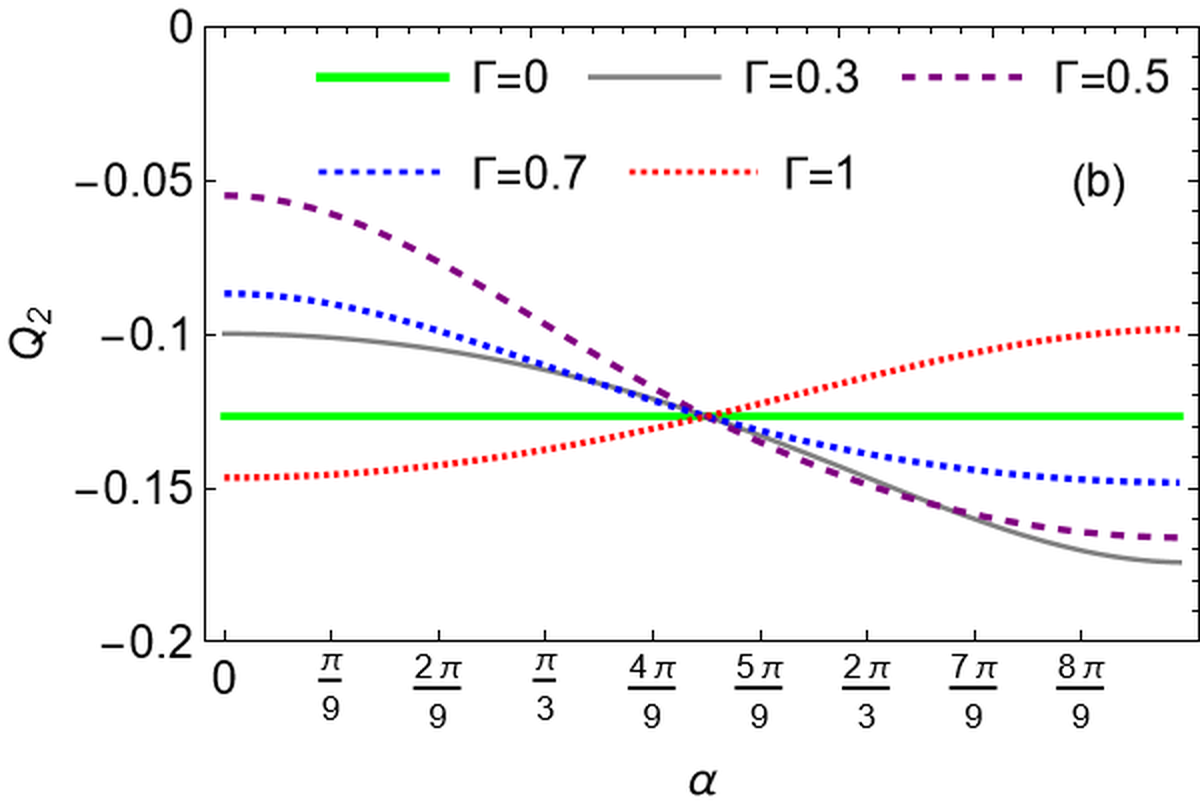}

\caption{\label{fig:3}Quadrature squeezing $Q_{2,\Psi}$: (a) $Q_{2,\Psi}$
as a function of the state parameter $\gamma$ for different coupling
strength parameters $\Gamma$, with the weak value fixed at $\langle\sigma_{x}\rangle_{w}=5.671$
corresponding to $\alpha=8\pi/9$. (b) $Q_{2,\Psi}$ as a function
of the weak value parameter $\alpha$ for different coupling strength
parameters $\Gamma$, with the state parameter $\gamma$ fixed at
10. Other parameters are the same as those in Fig. \ref{fig:2}.}
\end{figure}

To provide a clearer analysis of quadrature squeezing, we rely on
numerical calculations, with the results presented in Figs. \ref{fig:2}
and \ref{fig:3}. In Fig. \ref{fig:2}, we plot $Q_{1,\Psi}$ as a
function of the parameter $\gamma$ for various coupling strength
parameters $\Gamma$, while the weak value $\langle\sigma_{x}\rangle_{w}=5.671$
is fixed, corresponding to $\alpha=8\pi/9$. As shown in Fig. \ref{fig:2},
$Q_{1,\phi}$ ($\Gamma=0$) remains positive and increases with $\gamma$
{[}see the thick green curve{]}. Additionally, Fig. \ref{fig:2} demonstrates
that $Q_{1,\Psi}$ for cases where $\Gamma\neq0$ also consistently
takes positive values. This observation confirms that after the postselected
von Neumann measurement, the $F_{1}$ quadrature of the final MD state
$\vert\Psi\rangle$ still doesn't exhibit squeezing, consistent with
the behavior of the initial state.

In Fig. \ref{fig:3}, we present the numerical results for $Q_{2,\Psi}$,
which corresponds to the quadrature $F_{2}$ of the final MD state
$\vert\Psi\rangle$. In Fig. \ref{fig:3}(a), $Q_{2,\Psi}$ is plotted
as a function of the state parameter $\gamma$ for various coupling
strength parameters $\Gamma$, with a fixed large weak value $\langle\sigma_{x}\rangle_{w}=5.671$,
corresponding to $\alpha=8\pi/9$. Among these curves, the case $\Gamma=0$
(thick green curve) represents $Q_{2,\phi}$ for the initial state.
As shown, $Q_{2,\phi}$ decreases with increasing $\gamma$ and converges
to $-0.125$ across most parameter regions. Experimentally, the amount
of squeezing is quantified in terms of decibels (dB) \citep{Agarwal2013}.
This value of $-0.125$ corresponds to $9$ dB or $50\%$ squeezing.
Notably, in the specific case where the PND is $\delta=0$, $Q_{2,\phi}$
is consistently less than zero. Thus, the initial PCS $\vert\phi\rangle$
with $\delta=0$ always exhibits squeezing along the $F_{2}$ quadrature.

In Fig. \ref{fig:3}(a), after the postselected von Neumann measurements
($\Gamma\neq0$), we observe that the value of $Q_{2,\Psi}$ is lower
than $Q_{1,\phi}$ for the weak coupling strength parameter $\Gamma=0.3$.
This behavior persists over a larger range of the state parameter
$\gamma$, indicating that the squeezing characteristics of the quadrature
$F_{2}$ are enhanced under postselected von Neumann measurements.
However, as the coupling strength parameter $\Gamma$ increases, the
value of $Q_{2,\Psi}$ decreases further compared to the $\Gamma=0.3$
case. For $\Gamma=0.3$, the minimum value of $Q_{2,\Psi}$ reaches
approximately $-0.172$ near $\gamma=10$. This corresponds to a squeezing
level of $11$dB or $69\%$. Therefore, after performing postselected
WMs, the quadrature squeezing along the $F_{2}$ direction of the
field increases by approximately $19$\% under suitable system parameters
and weak values of the measured system observable. To further investigate
the signal amplification effect of weak values for weak system signals,
Fig. \ref{fig:3}(b) presents $Q_{2,\Psi}$ as a function of the weak
value angle $\alpha$ for different coupling strength parameters $\Gamma$
and a fixed $\gamma=10$. The numerical results in Fig. \ref{fig:3}(b)
indicate that the weak value amplifies the squeezing effects of $Q_{2,\Psi}$
when the coupling strength parameter $\Gamma$ is small, highlighting
the signal amplification capability of weak measurements. These results
demonstrate that the squeezing effects of $Q_{2,\Psi}$ can be enhanced
by postselected WMs.

\subsection{Sum squeezing }

The sum squeezing (SQ), a higher-order of squeezing, can also characterize
the nonclassicality of a two-mode radiation field. The concept of
two-mode SQ for a radiation field was introduced by Hillery \citep{squeezingPhysRevA.40.3147}.
The sum squeezing operator is defined as: 
\begin{equation}
V_{\varpi}=\frac{1}{2}\left(e^{i\varpi}a^{\dagger}b^{\dagger}+e^{-i\varpi}ab\right),\label{eq:sum squeezing operator}
\end{equation}
where the angle $\varpi\in\left[0,2\pi\right].$ Similar to single-mode
squeezing, the SQ degree $S_{ab}\left(\varpi\right)$ of a two-mode
radiation field is defined as 
\begin{equation}
S_{ab}\left(\varpi\right)=\frac{4\langle\left(\Delta V_{\varphi}\right)^{2}\rangle}{\langle N_{a}+N_{b}+1\rangle}-1,\label{eq:sum squeezing degree}
\end{equation}
where $\Delta V_{\varpi}^{2}=\langle V_{\varpi}^{2}\rangle-\langle V_{\varpi}\rangle^{2}$,
and particle number operators for the $a$ mode and $b$ mode are
$N_{a}=a^{\dagger}a$ and $N_{b}=b^{\dagger}b$, respectively. If
$-1\le S_{ab}\left(\varpi\right)<0$, the two-mode state exhibits
squeezing properties. The more negative the value of $S_{ab}\left(\varpi\right)$,
the greater the degree of sum squeezing in the state. The state is
maximally squeezed when $S_{ab}\left(\varpi\right)=-1$. By substituting
$V_{\varpi}$ into Eq. (\ref{eq:sum squeezing degree}), we obtain
the squeezing degree $S_{ab}\left(\varpi\right)$ of normal-ordering
operators: 
\begin{equation}
S_{ab}\left(\varpi\right)=\frac{2\left[Re[e^{-2i\varpi}\left\langle a^{2}b^{2}\right\rangle ]-2\left(Re[e^{-i\varpi}\langle ab\rangle]\right)^{2}+\left\langle N_{a}N_{b}\right\rangle \right]}{\left\langle N_{a}\right\rangle +\left\langle N_{b}\right\rangle +1}.\label{eq:sum queezing}
\end{equation}
The explicit expression of the SQ degree $S_{ab,\Psi}\left(\varpi\right)$
for the final MD state $\vert\Psi\rangle$ can be derived by substituting
the relevant expectation values, as listed in Appendix \ref{sec:A1}.
When $\Gamma=0$, no measurements is performed, and the state remains
as the initial PCS $\vert\phi\rangle$. For this trivial case, it
is straightforward to verify that the sum squeezing equals to zero,
i.e., $S_{ab,\phi}\left(\varpi\right)=0$. This result implies that,
regardless of the PND $\delta$ of the two-modes or the angle $\varpi$,
the SQ of the initial PCS $\vert\phi\rangle$ is always zero.

\begin{figure}
\raggedright
\includegraphics[width=8cm]{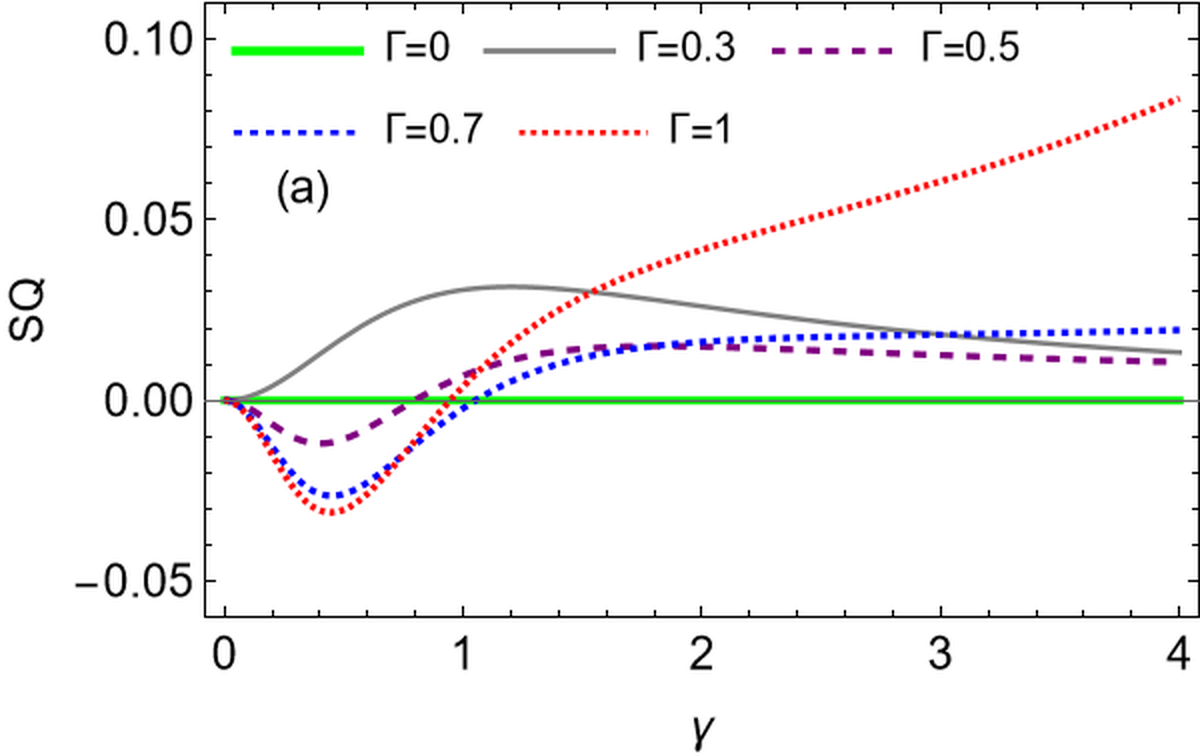}

\includegraphics[width=8cm]{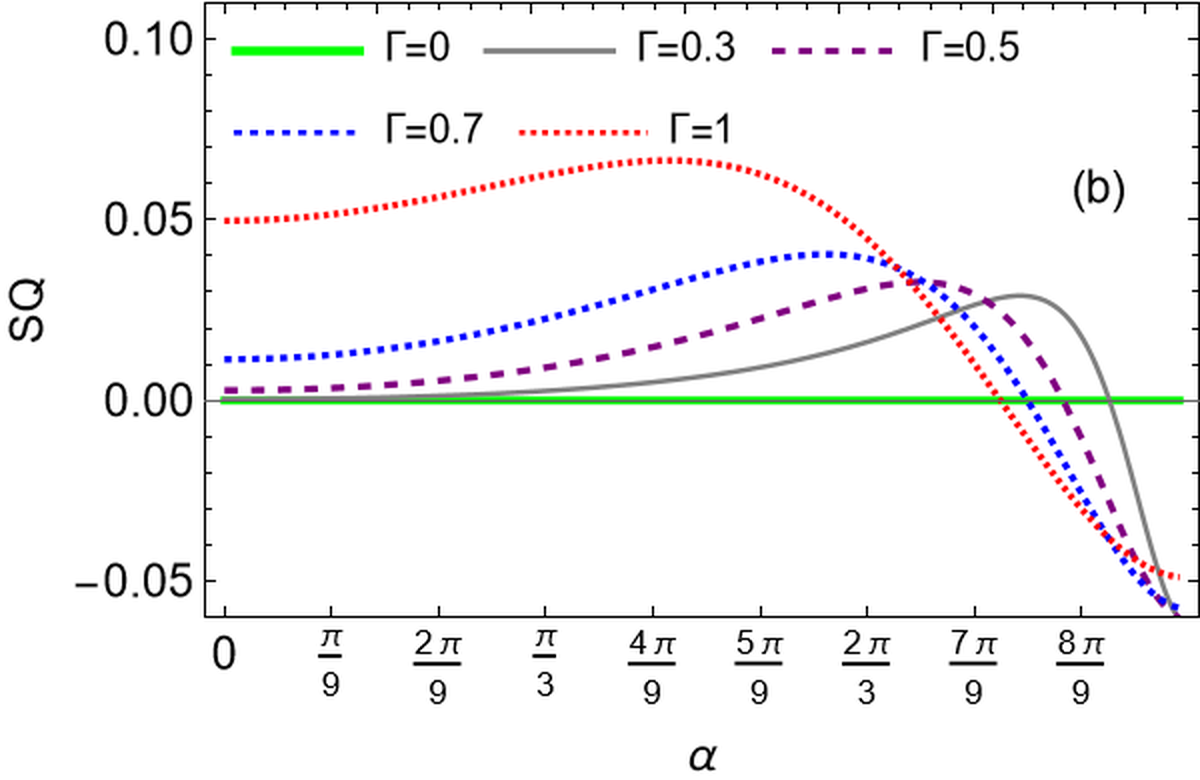}

\caption{\label{fig:4}The SQ of the final MD state $\vert\Psi\rangle$ under
postselected von Neumann measurement. (a) $S_{ab,\Psi}\left(\varpi\right)$
as a function of the state parameter $\gamma$ for different coupling
strength parameters $\Gamma$, while the weak vale is fixed to $\langle\sigma_{x}\rangle_{w}=5.671$.
(b) $S_{ab,\Psi}\left(\varpi\right)$ as a function of the weak value
parameter $\alpha$ for different coupling strength parameters $\Gamma$,
while the state parameter $\gamma$ is fixed at $0.5$. Here, we take
$\varpi=0$. Other parameters are the same as in Fig. \ref{fig:2}.}
\end{figure}

In Fig. \ref{fig:4}, we present the SQ degree $S_{ab,\Psi}\left(\varpi\right)$
for the final MD state $\vert\Psi\rangle$. In Fig. \ref{fig:4}(a),
we show $S_{ab,\Psi}\left(\varpi\right)$ as a function of the state
parameter $\gamma$ for different values of the coupling strength
parameter $\Gamma$, while fixing the weak value to $\langle\sigma_{x}\rangle_{w}=5.671$,
which corresponds to $\alpha=8\pi/9$. As shown in Fig. \ref{fig:4}(a),
SQ occurs in a small range of the state parameter $\gamma$ for appropriate
coupling strengths $\Gamma$, provided the weak value of the measured
system observable is large. For the anomalous weak value $\langle\sigma_{x}\rangle_{w}=5.671$,
that corresponds to $\alpha=8\pi/9$, the minimum value of SQ degree
$S_{ab,\Psi}\left(\varpi\right)$ occurs near $\gamma=0.5$ at larger
coupling strength parameters $\Gamma$. This squeezing effect diminishes
as $\gamma$ increases beyond this range. To further investigate the
dependence of SQ on the weak value, Fig. \ref{fig:4}(b) presents
the variation $S_{ab,\Psi}\left(\varpi\right)$ as a function of the
weak value parameter $\text{\ensuremath{\alpha}}$ for different values
of $\Gamma$, while keeping $\gamma=0.5$ fixed. As indicated in Fig.
\ref{fig:4}(b), the SQ degree $S_{ab,\Psi}\left(\varpi\right)$ takes
negative values only for anomalous weak values and non-zero coupling
strength parameters $\Gamma$. Notably, when the weak value angle
$\alpha$ approaches $\pi$, the SQ effect becomes more pronounced
for suitable coupling strength parameters $\Gamma$. This numerical
result further confirms the amplification effect of weak values in
postselected WMs within our scheme.

In the above discussions, we can conclude that the postselected von
Neumann measurement improves various squeezing characteristics of
our initial PCS $\vert\phi\rangle$. That is to say, after postselected
von Neumann measurement, the squeezing effects of $\vert\phi\rangle$
are enhanced for large anomalous weak values of the measured observable,
with proper composite state parameters $\gamma$ and coupling strength
parameters $\Gamma$. Besides the squeezing effect, quantum correlation
functions and entanglement, which exist between the two modes, also
characterize the nature of the PCS $\vert\phi\rangle$. In the next
sections, we investigate the effects of the postselected von Neumann
measurement on the quantum correlation functions and entanglement
features of the PCS $\vert\phi\rangle$.

\section{\label{sec:The-effects-quantum statics}The effects on quantum correlation
functions }

To examine the effects of postselected von-Neumann measurement on
the SOCC function $g_{ab}^{(2)}$ of the PCS, in this section, we
investigate the correlation functions between the two modes.

\subsection{Second-order cross-correlation function}

The normalized SOCC of the two-mode field is defined as \citep{PhysRevA.44.6043}
\begin{equation}
g_{ab}^{(2)}=\frac{\langle a^{\dagger}ab^{\dagger}b\rangle}{\langle a^{\dagger}a\rangle\langle b^{\dagger}b\rangle}.
\end{equation}
Here, $\langle a^{\dagger}ab^{\dagger}b\rangle$ represents the intensity-intensity
correlation between the two-modes, and $\langle a^{\dagger}a\rangle$
and $\langle b^{\dagger}b\rangle$ denote mean photon number for each
mode, respectively. This function characterize the correlation between
photons in the different modes. If $g_{a,b,\Psi}^{(2)}>1$, there
is correlation between the $a$-mode and $b$-mode of the two-mode
radiation field. Otherwise, they are inversely correlated. To discuss
the properties of $g_{a,b,\Psi}^{(2)}$, we first derive the average
values of $\langle a^{\dagger}ab^{\dagger}b\rangle$, $\langle a^{\dagger}a\rangle$,
and $\langle b^{\dagger}b\rangle$ under the state $\vert\Psi\rangle$.
Since their explicit expressions are cumbersome, we list them in Appendix
\ref{sec:A1}. In particular, when $\Gamma=0$, the SOCC $g_{ab,\Psi}^{(2)}$
reduces to $g_{ab,\phi}^{(2)}$ for the initial PCS $\vert\phi\rangle$,
and its expression can be written as 
\begin{equation}
g_{ab,\phi}^{(2)}=\frac{I_{\delta}(2|\gamma|){}^{2}}{I_{\delta-1}(2|\gamma|)I_{\delta+1}(2|\gamma|)}.\label{eq:28}
\end{equation}
As investigated in previous studies \citep{Agarwal:88,Lu}, the cross-correlation
function $g_{ab,\phi}^{(2)}$ between the $a$-mode and the $b$-mode
of the PCS $\vert\phi\rangle$ decreases as the PND $\delta$ and
the state parameter $\gamma$ increase, and eventually, $g_{ab,\phi}^{(2)}$
tends to a value of 1, indicating that the correlation between the
two modes is almost non-existent.

\begin{figure}
\includegraphics[width=8cm]{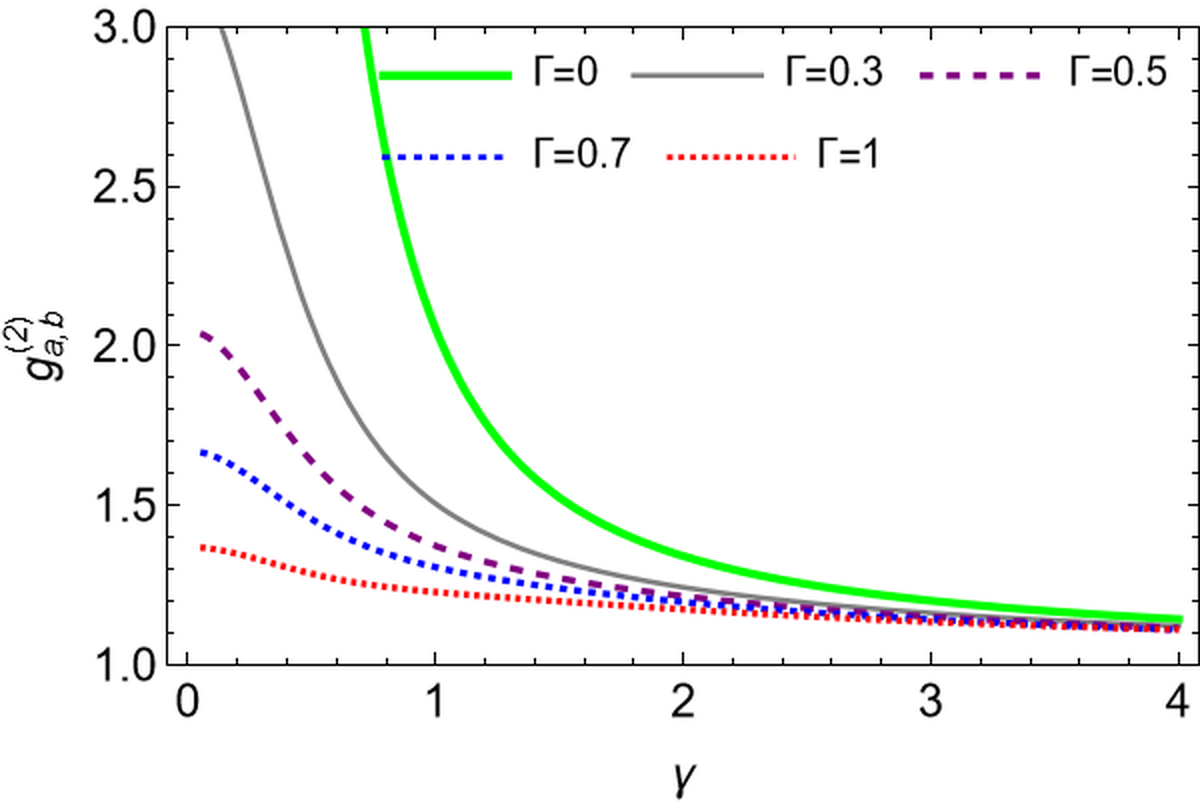}

\caption{\label{fig:5}SOCC $g_{ab,\Psi}^{(2)}(0)$ as a function of the state
parameter $\gamma$ for different coupling strength parameters $\Gamma$.
The green thick curve corresponds to the initial $\vert\phi\rangle$
case ($\Gamma=0$). Here, we take weak value as $\langle\sigma_{x}\rangle_{w}=5.761$
and the other parameters are the same as in Fig. \ref{fig:2}.}
\end{figure}

In Fig. \ref{fig:5} shows the variation of $g_{ab,\Psi}^{(2)}$ for
different system parameters. Specifically, in Fig. \ref{fig:5}, we
plot $g_{a,b,\Psi}^{(2)}$ as a function of the parameter $\gamma$
for different coupling strength parameters $\Gamma$, while fixing
the weak value to $\langle\sigma_{x}\rangle_{w}=5.761$. As shown
in Fig. \ref{fig:5}, the SOCC $g_{ab,\Psi}^{(2)}$ of the state $\vert\Psi\rangle$
decreases as the coupling strength parameter $\Gamma$ increases.
When the state parameter $\gamma$ is small, the value of $g_{ab,\Psi}^{(2)}(0)$
is greater than one. However, as $\gamma$ increases, all curves tend
to one. The higher the coupling strength parameters $\Gamma$, the
lower the correlation between the two modes of the final MD state
$\vert\Psi\rangle$. 

\subsection{second-order correlation function}

We can examine the statistical properties of the radiation field by
using the zero-time delay second-order correlation function $g^{(2)}(0)$.
This function measures the intensity correlation and serves as a crucial
parameter for characterizing the photon statistics of a source. For
a coherent source, $g^{(2)}(0)=1$, and for a single-photon state
$g^{(2)}(0)=0$ \citep{RN164}. If $0\le g^{(2)}(0)<1$, the field
exhibits sub-Poissonian statistics, characterizing the non-classical
features of the associated radiation field \citep{Agarwal2013}. For
our state $\vert\Psi\rangle$, the function $g^{(2)}(0)$ for each
mode can be written as: 

\begin{align}
g_{a}^{(2)} & =\frac{\langle a^{\dagger2}a^{2}\rangle}{\langle a^{\dagger}a\rangle},\\
g_{b}^{(2)} & =\frac{\langle b^{\dagger2}b^{2}\rangle}{\langle b^{\dagger}b\rangle}.
\end{align}
We derived the explicit expressions of $g^{(2)}(0)$ by calculating
the average values of the associated operators (see Appendix \ref{sec:A1}).
Figure \ref{fig:7} displays the variations of $g_{a,\Psi}^{(2)}(0)$
and $g_{b,\Psi}^{(2)}(0)$ with the state parameter $\gamma$ for
different coupling strength parameters $\Gamma$. We fixed the weak
value parameter at $\alpha=8\pi/9$ to analyze the effects of large
anomalous weak values on the statistical properties of the final MD
state $\vert\Psi\rangle$. As shown in Fig. \ref{fig:7}, the $\Gamma=0$
case corresponds to the $g^{(2)}(0)$ of the initial PCS state $\vert\phi\rangle$.
Its value for both $a$ and $b$-modes is smaller than one and approaches
one as the state parameter $\gamma$ increases {(}see green curves
in Figs. \ref{fig:7}(a) and \ref{fig:7}(b){)}. After the postselected
von Neumann measurement ($\Gamma\neq0$), for small parameter $\gamma$,
the values of the second-order correlation functions $g_{a}^{(2)}(0)$
and $g_{b}^{(2)}(0)$ quickly reach their minimum values with increasing
coupling strength parameter $\Gamma$, similar to the single-photon
state case. On the other hand, for large values of the state parameter
($\gamma\gg1$), $g_{a}^{(2)}(0)$ and $g_{b}^{(2)}(0)$ become indistinguishable
for all coupling strength parameters $\Gamma$ and tend to one, which
corresponds to the coherent state case. This result also explains
why the SOCC $g_{ab,\Psi}^{(2)}$ of $\vert\Psi\rangle$ approaches
one when $\gamma\gg1$ {(}see Fig. \ref{fig:5}{)}.

It is interesting to note that when comparing the green thick curve
($\Gamma=0$) with other curves in Figs. \ref{fig:7}(a) and \ref{fig:7}(b),
the second-order correlation function of the $a$-mode and $b$-mode
changes more dramatically after the postselected measurement ($\Gamma\neq0$).
Due to the weak value amplification effects, this result allows us
to achieve the characteristics of a single-photon state in small regions
of the state parameter $\gamma$. Our numerical results suggest that
following the postselected von Neumann measurement, the two modes
of the final MD state $\vert\Psi\rangle$ could potentially generate
a single-photon field \citep{RevModPhys.54.1061,Lounis_2005}. 

\begin{figure}
\includegraphics[width=8cm]{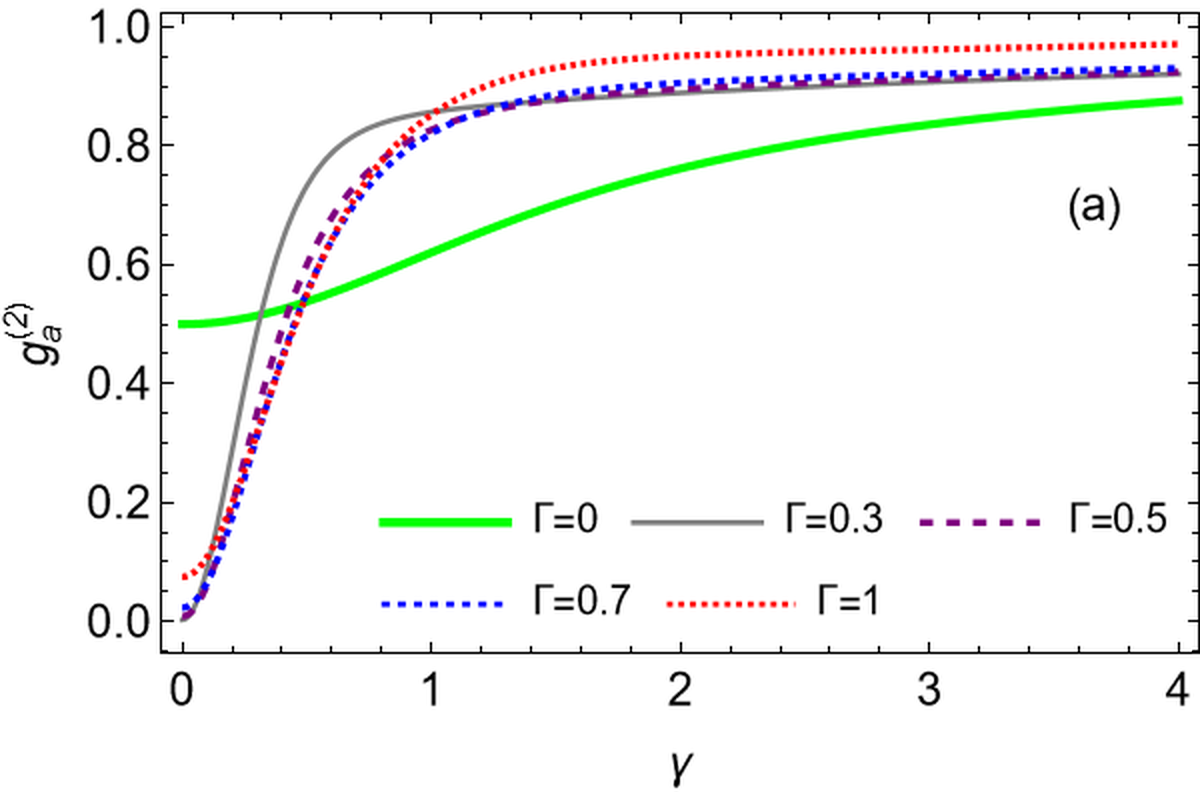}

\includegraphics[width=8cm]{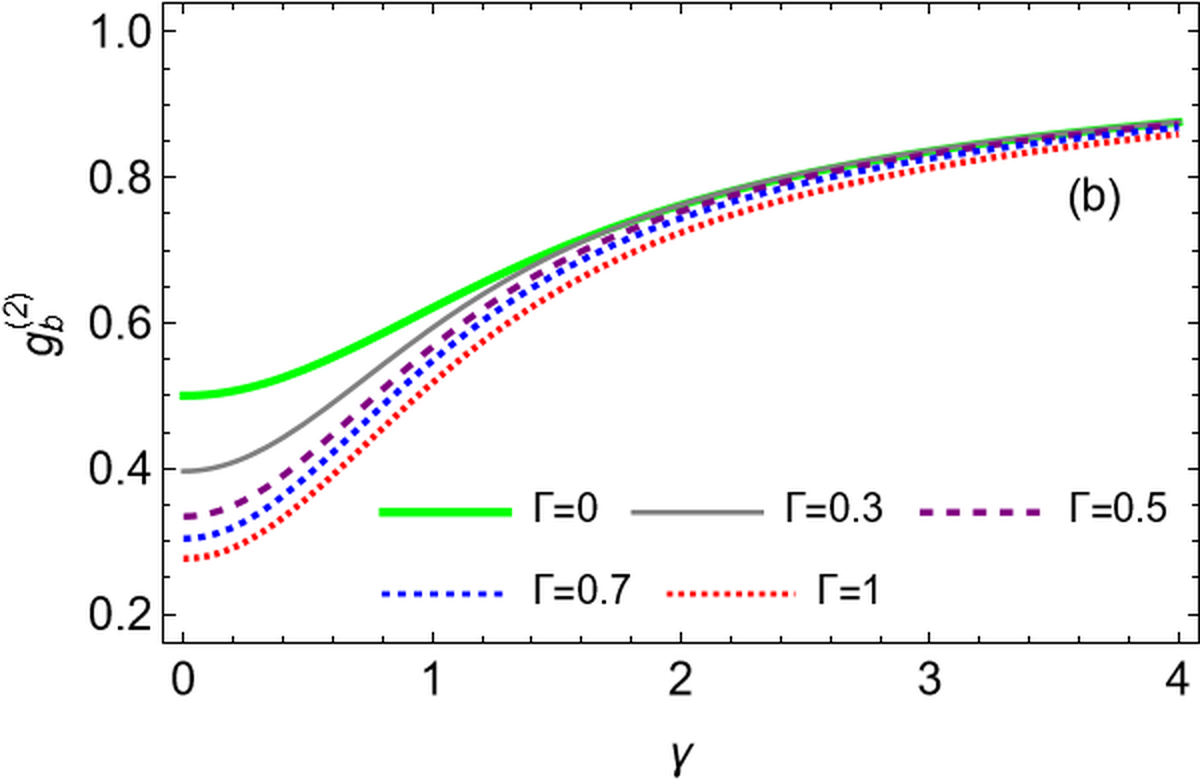}

\caption{\label{fig:7}Second-order correlation as a function of the $a$ and
$b$ modes of the final MD state $\vert\Psi\rangle$. (a) $g_{a,\Psi}^{(2)}(0)$
as a function of the state parameter $\gamma$ for different coupling
strength parameters $\Gamma$ for mode $a$. (b) the $b$ mode's $g_{b,\Psi}^{(2)}(0)$
as a function of the state parameter $\gamma$ for different coupling
strength parameters $\Gamma$ . Here, we take weak value as $\langle\sigma_{x}\rangle_{w}=5.761$
and other parameters are the same as in Fig. \ref{fig:2}.}
\end{figure}

\section{\label{sec:The-effects entanglement}The effects on entanglement}

In this section, we investigate the effects of postselected von Neumann
measurement on quantum correlations of the PCS. In a previous study
\citep{Agarwal_2005}, the inseparability of the PCS were confirmed
in light of the Peres-Horodecki criterion and various entropies. Here,
we separately examine the HZ correlation and EPR correlation to analyze
the inseparability of the final MD state $\vert\Psi\rangle$, as defined
in Eq. (\ref{eq;postselection}). 

\subsection{HZ correlation }

\begin{figure}
\includegraphics[width=8cm]{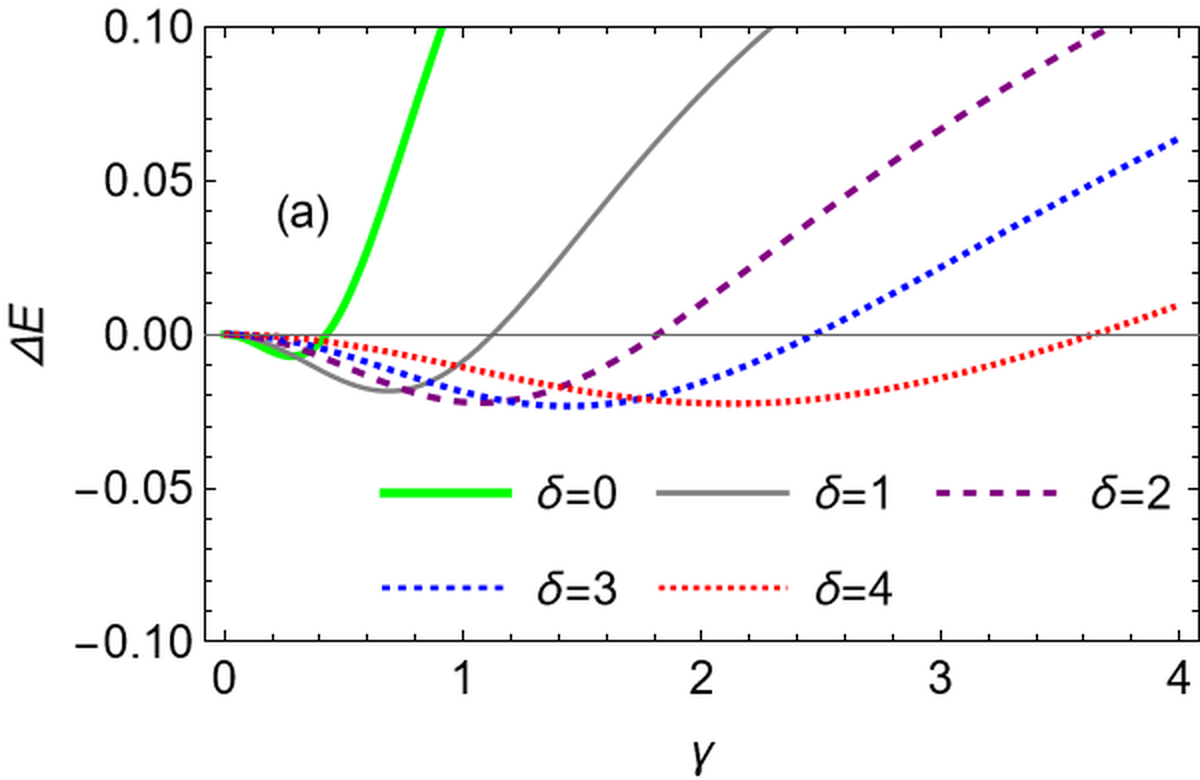}

\includegraphics[width=8cm]{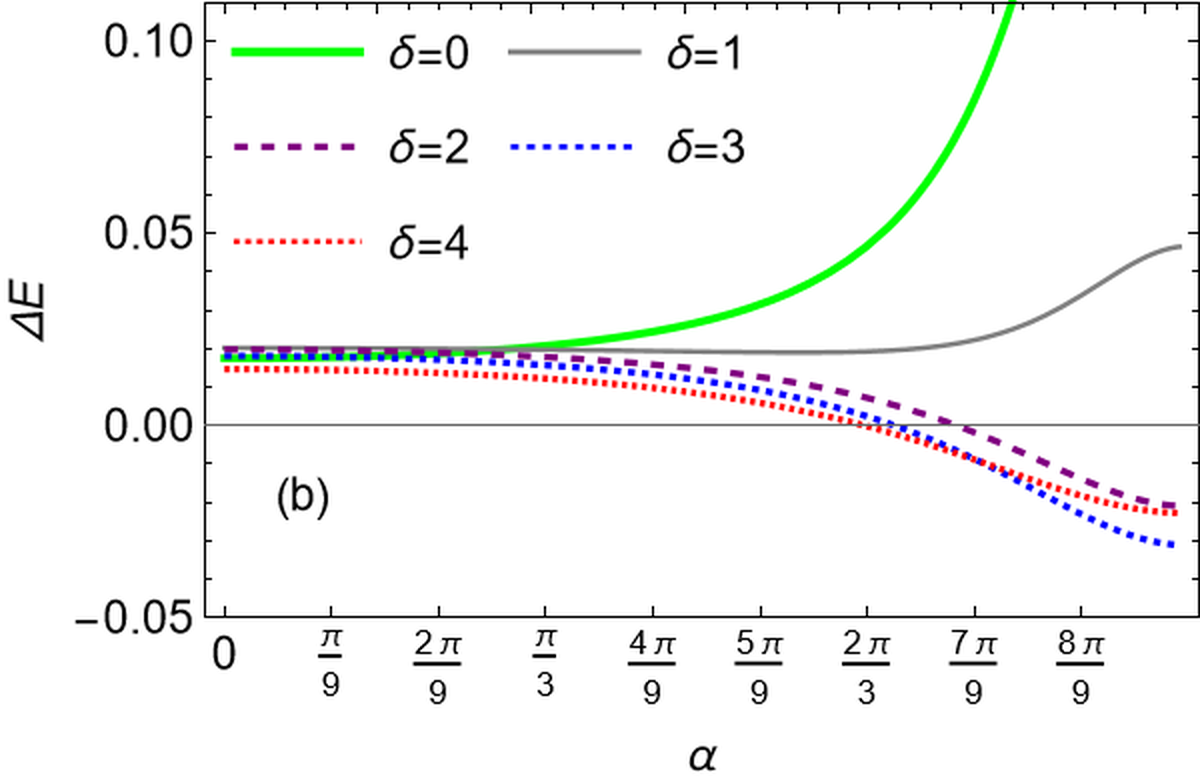}

\caption{\label{fig:7-1}HZ correlation between two modes of the final MD state
$\vert\Psi\rangle$. (a) $\Delta E$ as a function of the state parameter
$\gamma$ for different PND values $\delta$, with $\langle\sigma_{x}\rangle_{w}=5.761$.
(b) $\Delta E$ as a function of the weak value parameter $\alpha$
for different PND values $\delta$, with $\gamma=1.5$. Here, we take
$\Gamma=0.3$ and other parameters are the same as in Fig. \ref{fig:2}.}

\end{figure}

First, we study the entanglement between the two modes in the light
field by adopting the inequality provided by Hillery and Zubairy \citep{EntanglementTwo-Mode}.
This is defined as \citep{Daoming2015QuantumPO}:
\begin{equation}
E=\langle N_{a}\rangle\langle N_{b}\rangle-|\langle ab\rangle|^{2}.\label{eq:31}
\end{equation}
If $E<0$, it implies that the state exhibits entanglement between
the two modes. From the CSI, $|\langle ab\rangle\vert^{2}\le\langle a^{\dagger}a\rangle\langle bb^{\dagger}\rangle$
\citep{PhysRevResearch.3.033095}, the entanglement condition for
two-mode fields characterized by HZ correlation is bounded as $-\langle N_{a}\rangle\le E<0$.
It is straightforward to verify that the HZ correlation function $E$
for the initial PCS $\vert\phi\rangle$ can expressed as: 
\begin{align}
E_{\phi} & =\frac{|\gamma|^{2}I_{\delta-1}(2|\gamma|)I_{\delta+1}(2|\gamma|)}{I_{\delta}(2|\gamma|){}^{2}}-|\gamma|^{2}\nonumber \\
 & =|\gamma|^{2}\left(\frac{1}{g_{ab,\phi}^{(2)}}-1\right),\label{eq:32}
\end{align}
where $g_{ab,\phi}^{(2)}$ is the SOCC defined in Eq. \ref{eq:28}.
To evaluate the HZ correlation for our output state $|\Psi\rangle$,
we compute $E_{\Psi}$ using Eq. (\ref{eq:31}), substituting the
average values of $\langle ab\rangle,\,\langle a^{\dagger}a\rangle$and
$\langle b^{\dagger}b\rangle$ as listed in Appendix \ref{sec:A1}.
To demonstrate the effects of postselected von Neumann measurement
on the entanglement of the output state $|\Psi\rangle$, we introduce
the difference between $E_{\Psi}$ and $E_{\phi}$, defined as $\Delta E=E_{\Psi}-E_{\phi}$.
Fig. \ref{fig:7-1} presents the numerical results for $\Delta E$.
If $\Delta E<0$, it indicates that the degree of entanglement has
increased compared to the initial state; otherwise, it has decreased.
In Fig. \ref{fig:7-1}(a), $\Delta E$ is plotted as a function of
the state parameter $\gamma$ for different values of PND denoted
by $\delta$, with the coupling strength parameter $\Gamma$ fixed
in the WM regime ($\Gamma=0.3$) and a large anomalous weak value
$\langle\sigma_{x}\rangle_{w}=5.761$. The results show that $\Delta E$
takes negative values for certain regions of $\gamma$, and these
regions expand as $\delta$ increases. This result indicates that
the final MD state $\vert\Psi\rangle$ exhibits stronger entangled
properties than the initial PCS $\vert\phi\rangle$ in appropriate
parameter regions. To further validate the role of weak value amplification
for enhancing the entanglement between two modes of PCS after postselected
measurement, Fig. \ref{fig:7-1}(b) illustrates the variation of $\Delta E$
with the weak value parameter $\alpha$ for different values of $\delta$.
The curves in Fig. \ref{fig:7-1}(b) reveal that, within the WM regime,
if the PND $\delta$ between modes $a$ and $b$ of the PCS is equal
to or greater than two (i.e., $\delta\ge2$), the entanglement between
the two modes of the measurement output state $\vert\Psi\rangle$
surpasses that of the initial state $\vert\phi\rangle$ for large
anomalous weak values.

\subsection{EPRs correlation}

Another inseparability criterion for two-mode systems in multimode
continuous variable systems relies on the total variance of a pair
of EPR-type operators \citep{PhysRevLett.84.2722}. This criterion
provides a sufficient condition for entanglement in any two-party
continuous variable state and is also a necessary and sufficient condition
for inseparability. For a two-mode system, the EPR correlation is
defined in terms of the variances of the EPR-type operators $X_{1}-X_{2}$
and $P_{1}-P_{2}$ as \citep{REN2019106} 

\begin{eqnarray}
I & = & \langle\Delta^{2}\left(X_{1}-X_{2}\right)\rangle+\langle\Delta^{2}\left(P_{1}-P_{2}\right)\rangle\nonumber \\
 & = & 2\left(1+\langle a^{\dagger}a\rangle+\langle b^{\dagger}b\rangle-\langle a^{\dagger}b^{\dagger}\rangle-\langle ab\rangle\right)\nonumber \\
 & - & 2\left(\langle a\rangle-\langle b^{\dagger}\rangle\right)\left(\langle a^{\dagger}\rangle-\langle b\rangle\right).\label{eq:ERP}
\end{eqnarray}
where $X_{1}=\frac{a+a^{\dagger}}{\sqrt{2}},\;X_{2}=\frac{b+b^{\dagger}}{\sqrt{2}}$,
$P_{1}=\frac{a-a^{\dagger}}{i\sqrt{2}}$ and $P_{2}=\frac{b-b^{\dagger}}{i\sqrt{2}}$.
If the total variance is less than 2, $I<2$, the two-mode system
can be considered inseparable, indicating quantum entanglement. Otherwise,
the system is classical. The smaller the value of $I$, the stronger
the EPR correlation.

\begin{figure}
\includegraphics[width=8cm]{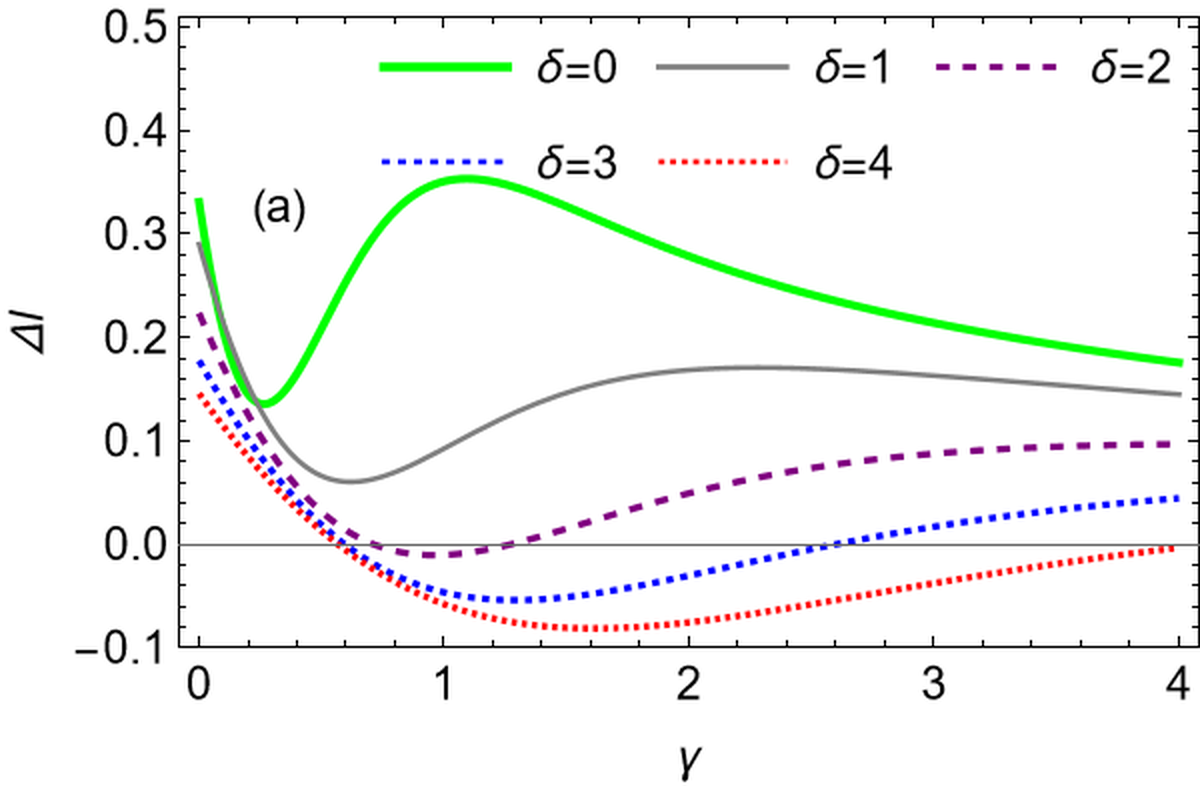}

\includegraphics[width=8cm]{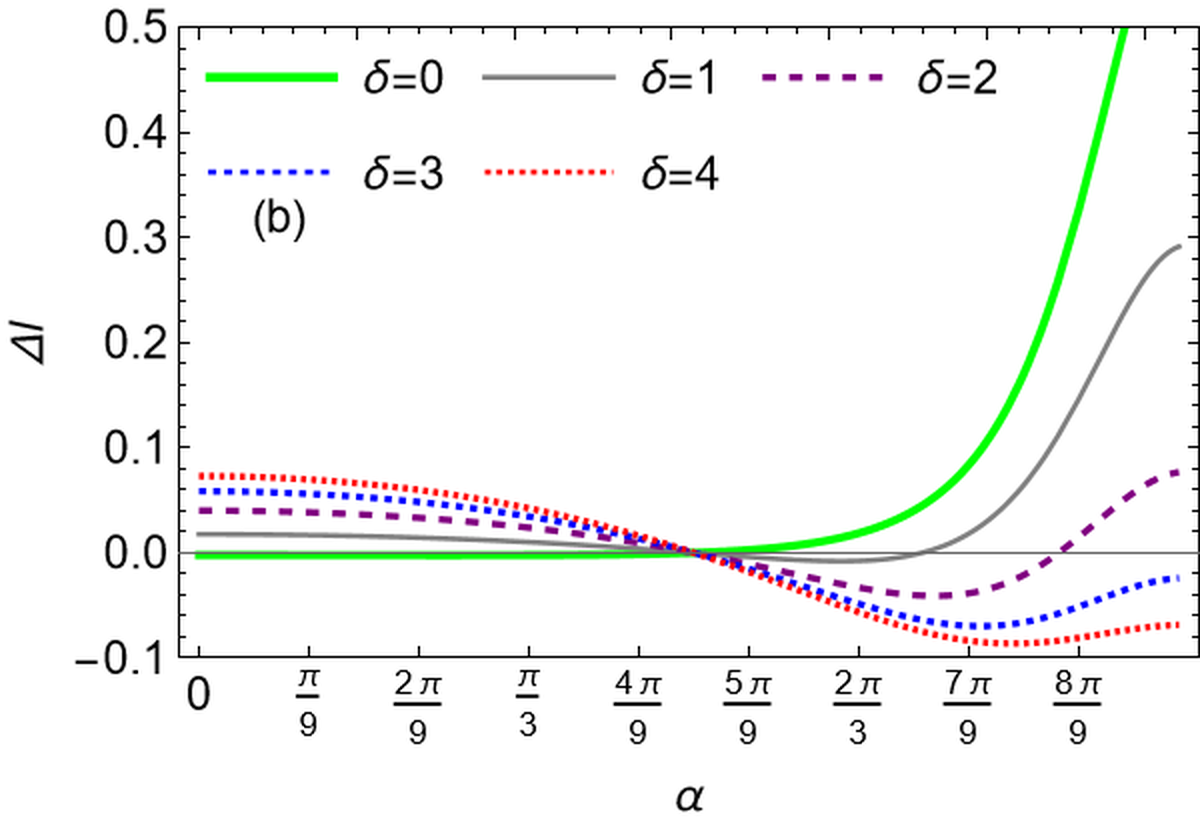}

\caption{\label{fig:8}EPR correlation differences $\Delta I$ for the final
MD state $\vert\Psi\rangle$ and initial PCS $\vert\phi\rangle$.
(a) $\Delta I$ as a function of the state parameter $\gamma$ for
different PND $\delta$ between the two modes of the state $\vert\phi\rangle$.
(b) $\Delta I$ as a function of the weak value angle $\alpha$ for
different PND $\delta$ between the two modes of the state $\vert\phi\rangle$.
Here $\Gamma=0.3$, with (a) $\langle\sigma_{x}\rangle_{w}=5.761$,
and (b) $\gamma=1.5$. Other parameters are the same as in Fig. \ref{fig:2}.}

\end{figure}

The explicit expression of the EPR correlation $I_{\Psi}$ of our
postselected measurement output state $\vert\Psi\rangle$ can obtained
by substituting the associated averages listed in Appendix \ref{sec:A1}
into Eq. (\ref{eq:ERP}). As as special case, when $\Gamma=0$, $I_{\Psi}$
reduces to the EPR correlation $I_{\phi}$ of the initial PCS $\vert\phi\rangle$,
and its expression reads as 
\begin{align}
I_{\phi} & =2\left(\frac{|\gamma|I_{\delta-1}(2|\gamma|)}{I_{\delta}(2|\gamma|)}+\frac{|\gamma|I_{\delta+1}(2|\gamma|)}{I_{\delta}(2|\gamma|)}-2\Re(\gamma)+1\right)\nonumber \\
 & =8Q_{2,\phi}+2=8\Delta F_{2,\phi}^{2}.\label{eq:34}
\end{align}
Here, $Q_{2,\phi}$ is the quadrature squeezing parameter of the initial
PCS $\vert\phi\rangle$ as defined in Eq. (\ref{eq:23}), and $\Delta F_{2,\phi}^{2}$
represents the variance of the quadrature operator $F_{2}$, defined
in Eq. (\ref{eq:17}), over the state $\vert\phi\rangle$. For the
PCS state $\vert\phi\rangle$, its EPR correlation is proportional
to the variance $\Delta F_{2,\phi}^{2}$. To analyze the effects of
the postselected von Neumann measurement on the EPR correlation between
the two modes of the state $\vert\Psi\rangle$, we introduce the difference
between $I_{\Psi}$ and $I_{\phi}$, defined as $\Delta I=I_{\Psi}-I_{\phi}$.
To observe these effects, we plotted $\Delta I$ as a function of
various system parameters, with the numerical results shown in Fig.
\ref{fig:8}. The variation of $\Delta I$ with changes in the state
parameter $\gamma$ for different PND $\delta$ is presented in Fig.
\ref{fig:8}(a). When the weak value is set to $\langle\sigma_{x}\rangle_{w}=5.761$
and the coupling strength parameter is fixed at $\Gamma=0.3$, $\Delta I$
takes negative values in a range of the state parameter $\gamma$
for cases where $\delta\ge2$. The regions of negative values become
broader and deeper as the PND $\delta$ between the two modes increases.
This result implies that in the postselected WM regime ($\Gamma<1$)
with large anomalous weak values, the EPR correlation between the
two modes of the final MD state $\vert\Psi\rangle$ is stronger than
that of the initial PCS state $\vert\phi\rangle$ when $\delta\ge2$.
Similar to the HZ correlation case, we examined the dependence of
the enhancement of the entanglement on the anomalous weak values.
Fig. \ref{fig:8}(b) shows $\Delta I$ as a function of the angle
$\alpha$. The curves in Fig. \ref{fig:8}(b) show that larger PND
$\delta$ does not enhance the violation of the EPR correlation for
small weak value angles $\alpha$. However, for larger weak value
angles $\alpha$, $\Delta I$ becomes negative, and its absolute value
increases with increasing PND $\delta$.

In summary, the large anomalous weak values $\langle\sigma_{x}\rangle_{w}$
of the measured system observable contribute significantly to the
weak signal amplification observed in our scheme.

\section{\label{sec:6}Joint Wigner function}

To gain a deeper understanding of the effects of postselected von
Neumann measurements on the properties of $\vert\Psi\rangle$, we
examine the phase space distribution by calculating its Wigner function.
The Wigner function of a single-mode radiation field can be written
as \citep{Agarwal2013}:
\begin{equation}
W\left(\alpha\right)=\frac{2}{\pi}Tr\left[\rho D(\alpha)PD^{\dagger}(\alpha)\right],\label{eq:35-1}
\end{equation}
where $\rho$ is the density operator of the given state, and $P=e^{i\pi a^{\dagger}a}$
is the parity operator. However, in our current work, this Wigner
function cannot be used directly since the PCS is a two-mode radiation
field. In the two-mode case, the Wigner function of the final MD state
$\vert\Psi\rangle$ can be expressed in terms of the joint Wigner
function
\begin{align}
W_{J}\left(\alpha,\beta\right) & =\frac{4}{\pi^{2}}Tr\left[\rho D(\alpha)D(\beta)P_{j}D^{\dagger}(\beta)D^{\dagger}(\alpha)\right]\nonumber \\
 & =\frac{4}{\pi^{2}}P_{J}(\alpha,\beta),\label{eq:36-1}
\end{align}
where $\rho=\vert\Psi\rangle\langle\Psi\vert$ is the density operator
of the state $\vert\Psi\rangle$, $P_{j}=P_{a}P_{b}=e^{i\pi a^{\dagger}a}e^{i\pi b^{\dagger}b}$
is the joint photon number parity operator, and $D(\alpha)=e^{\alpha a^{\dagger}-\alpha^{\ast}a}$
and $D(\beta)=e^{\beta b^{\dagger}-\beta^{\ast}b}$ are the displacement
operators acting on the $a$ and $b$ modes of the two-mode state
$\vert\Psi\rangle$, respectively. The joint Wigner function $W_{J}$
is a function in the $4D$ phase space, whose coordinates are {(}$Re[\alpha]$,
$Im[\alpha]$, $Re[\beta]$, $Im[\beta]${)}. In Eq. (\ref{eq:36-1}),
$P_{J}(\alpha,\beta)$ is referred to as the scaled Wigner function,
which we use for our analysis of the phase space distribution of the
measurement output state $\vert\Psi\rangle$. Upon examining the expression
for $P_{J}(a,\beta)$, we find that it is the average value of the
joint parity operator $P_{j}$ after applying displacement operators
with amplitudes $-\alpha$ and $-\beta$ to the $a$ and $b$ modes,
respectively. Since the eigenstates of the parity operators $P_{a}$
and $P_{b}$ are the Fock states $\vert n\rangle_{a}$ and $\vert n\rangle_{b}$,
with eigenvalues $(-1)^{n_{a}}$ and $(-1)^{n_{b}}$, the scaled Wigner
function is therefore bounded by $\pm1$, i.e., 
\begin{equation}
-1\le P_{J}(\alpha,\beta)\le1.\label{eq:37-1}
\end{equation}

\begin{figure}
\includegraphics[width=8cm]{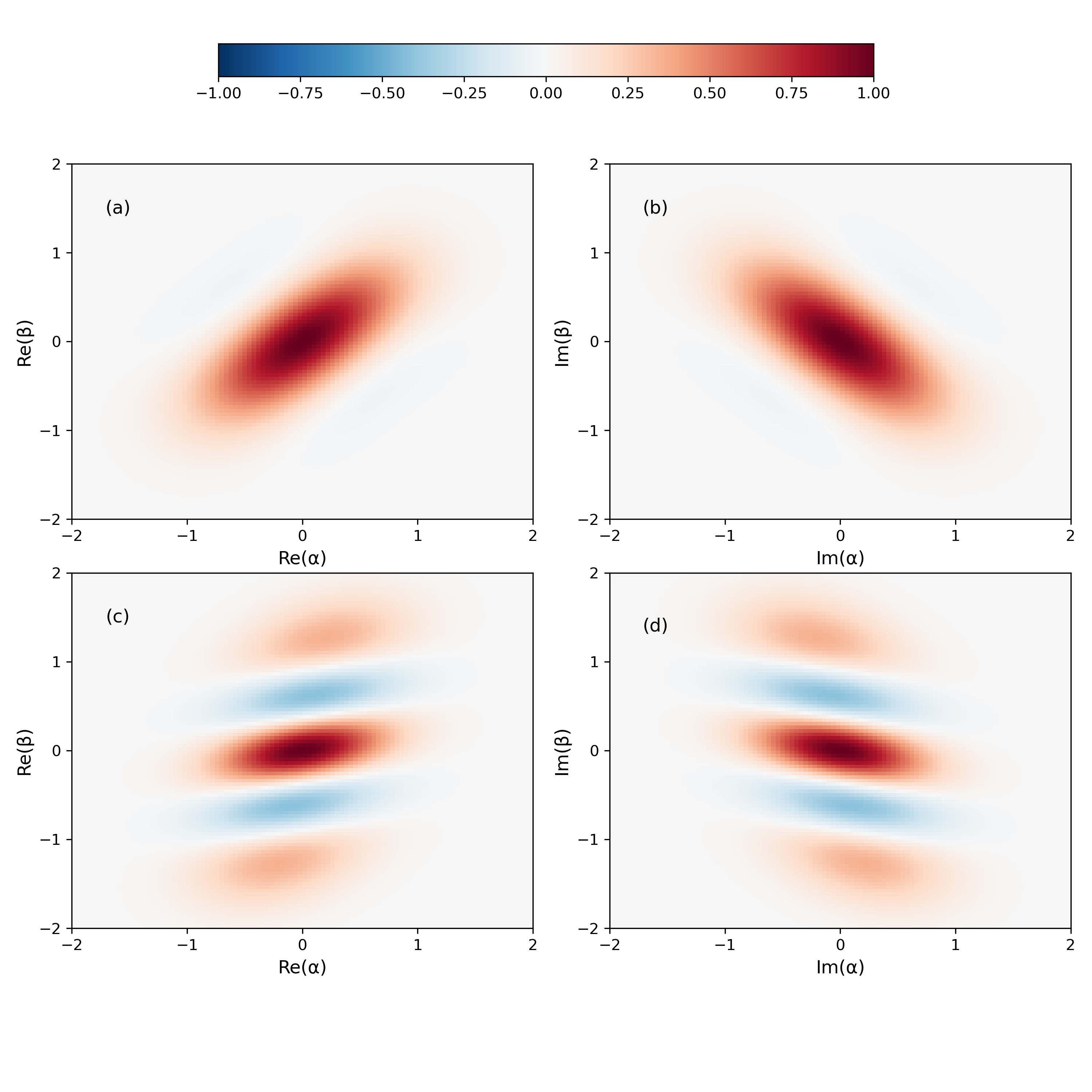}

\caption{\label{fig:10}Cuts in the scaled two-mode Wigner function. (a and
b) A 2D plane cut along (a) the $Re(\alpha)-Re(\beta)$ axes and (b)
$Im(\alpha)-Im(\beta)$ axes of the calculated $4D$ scaled joint
Wigner function $P_{J}(\alpha,\beta)$ for the initial PCS state $\vert\phi\rangle$
with $\delta=0$. (c and d) A 2D plane cut along (c) the $Re(\alpha)-Re(\beta)$
axes and (d) the $Im(\alpha)-Im(\beta)$ axes of the calculated $4D$
scaled joint Wigner function $P_{J}(\alpha,\beta)$ of the initial
PCS state $\vert\phi\rangle$ with $\delta=2$. Here, $\gamma=0.5$.}
\end{figure}

\begin{figure}
\includegraphics[width=8cm]{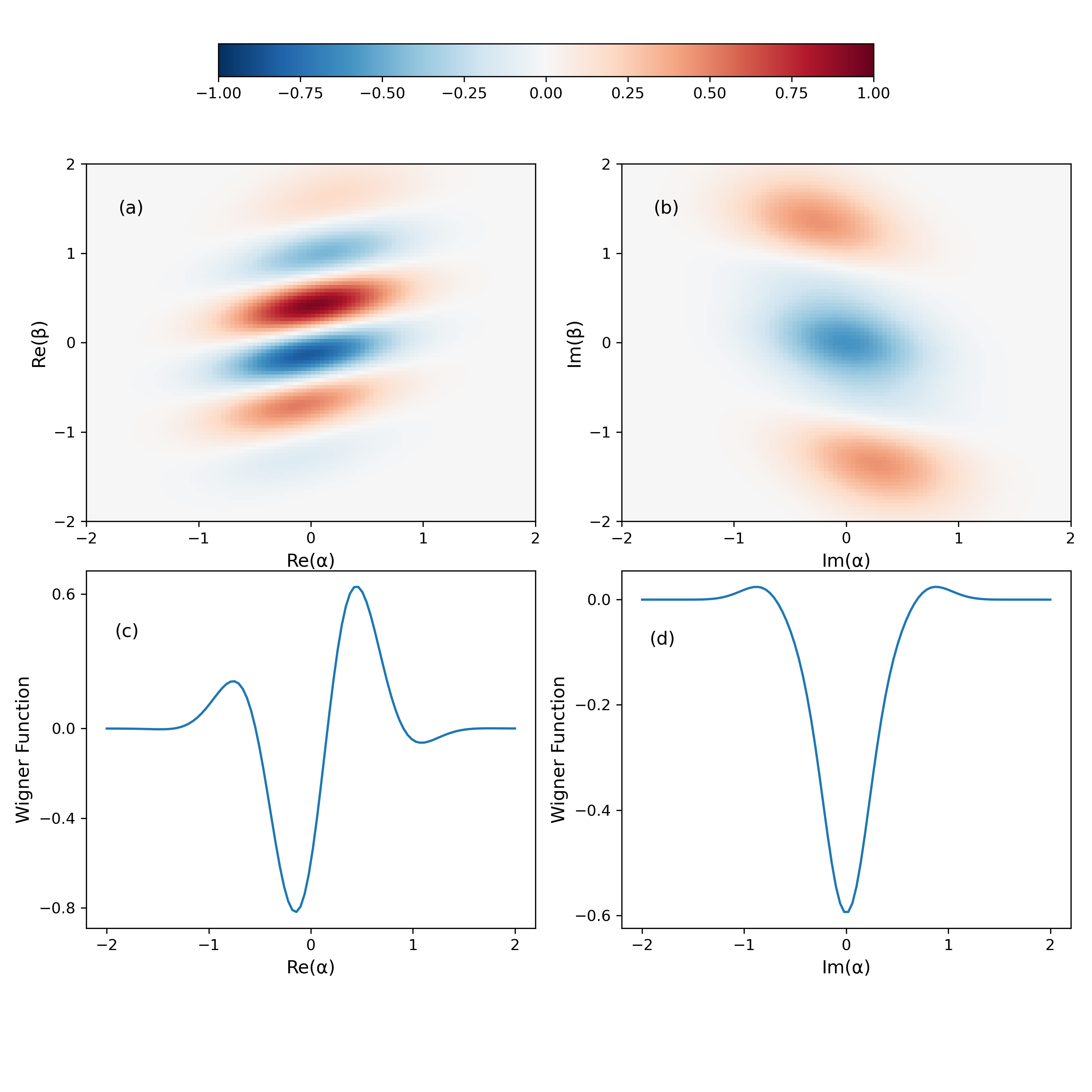}

\caption{\label{fig:11-1}Joint Wigner function of the final MD state $\vert\Psi\rangle$
after postselected measurement.(a and b) A 2D plane cut along (a)
the $Re[\alpha]-Re[\beta]$ axes and (b) the $Im[\alpha]-Im[\beta]$
axes of the calculated $4D$ scaled joint Wigner function $P_{J}(\alpha,\beta)$
for the measurement output state $\vert\Psi\rangle$ with $\Gamma=0.3$,
$\gamma=0.5$, $\langle\sigma_{x}\rangle_{w}=5.761$ and $\delta=2$.
(c) Diagonal line cuts of the data shown in (a), corresponding to
a $1D$ plot of the calculated scaled joint Wigner function $P_{J}(\alpha,\beta)$
along $Re[\alpha]=Re[\beta]$ with $Im[\alpha]=Im[\beta]=0$. (d)
Diagonal line cuts of the data shown in (b), corresponding to a $1D$
plot of the calculated scaled joint Wigner function $P_{J}(\alpha,\beta)$
along $Im[\alpha]=Im[\beta]$ with $Re[\alpha]=Re[\beta]=0$.}
\end{figure}
To illustrate the core features in this 4D Wigner function of the
final MD state $\vert\Psi\rangle$ and compare it with the initial
PCS state $\vert\phi\rangle$, we first present the 2D cuts along
the $Re[\alpha]-Re[\beta]$ plane and the $Im[\alpha]-Im[\beta]$
plane of the joint Wigner function of $\vert\phi\rangle$, as shown
in Fig. \ref{fig:10}. Figs. \ref{fig:10}(a) and \ref{fig:10}(b)
represent the density plot of the joint Wigner function of the initial
PCS state $\vert\phi\rangle$ with $\gamma=0.5$ and $\delta=0$,
in the planes $Im[\alpha]=Im[\beta]=0$ and $Re[\alpha]=Re[\beta]=0$,
respectively. As shown in these figures, the initial PCS $\vert\phi\rangle$
is Gaussian when there is no PND between the two modes ($\delta=0$),
and it exhibits squeezing along the $F_{2}$ quadrature {[}see Sec.\ref{sec:The-effects squeezing}
A{]}. Similarly, Figs. \ref{fig:10}(c) and \ref{fig:10}(d) show
the density plots of the joint Wigner function of the initial PCS
state $\vert\phi\rangle$ with $\gamma=0.5$ and $\delta=2$, in the
planes $Im[\alpha]=Im[\beta]=0$ and $Re[\alpha]=Re[\beta]=0$, respectively.
It is evident that when there is PND between the two modes ($\delta=2$),
the state $\vert\phi\rangle$ becomes non-Gaussian. Additionally,
the characteristic interference fringes of the PCS are visible, indicating
substantial coherence and consistent multiphoton phases. This result
suggests that PND ($\delta\neq0$) between the two modes is crucial
for enhancing the nonclassicality of the PCS state $\vert\phi\rangle$.

In Fig. \ref{fig:11-1}, we present the scaled joint Wigner function
$P_{J}(\alpha,\beta)$ of the final MD state $\vert\Psi\rangle$ with
$\Gamma=0.3$, $\gamma=0.5$, $\langle\sigma_{x}\rangle_{w}=5.761$,
and $\delta=2$. We observe that this Wigner function exhibits highly
nonclassical characteristics in phase space. Compared to the initial
PCS state $\vert\phi\rangle$ {[}see Figs. \ref{fig:10} (c) and \ref{fig:10}
(d){]}, we can see that after the postselected measurement, the shapes
of the scaled joint Wigner functions not only show good squeezing
for the appropriate coupling strength parameter $\Gamma$ and large
anomalous weak value $\langle\sigma_{x}\rangle_{w}$, but we also
can observe clear quantum interference structures formed between the
peaks {[}see Figs. \ref{fig:11-1} (c) and \ref{fig:11-1} (d){]}.
Furthermore, if someone take comparison Fig. \ref{fig:10} and Fig.
\ref{fig:11-1} who can find that after postselected von Neumann measurement
the negative values of Wigner function increased for large anomalous
weak value compared to initial state case. This indicated that the
nonclassicality of the PCS enhanced after the postselected von Neumann
measurement. 

These phase space analyses demonstrate that after the postselected
von Neumann measurement, the nonclassicality and non-Gaussianity of
PCS state $\vert\phi\rangle$ is enhanced due to the weak signal amplification
feature of the weak value.

\section{\label{sec:Fidelity}Fidelity }

To deeply understand the effects of postselected von Neumann measurement
on the initial PCS, in this section we study state distance between
initial and final MD state corresponding to the before and after measurement
process, and the teleportation fidelity. 

\subsection{State distance }

\begin{figure}
\includegraphics[width=8cm]{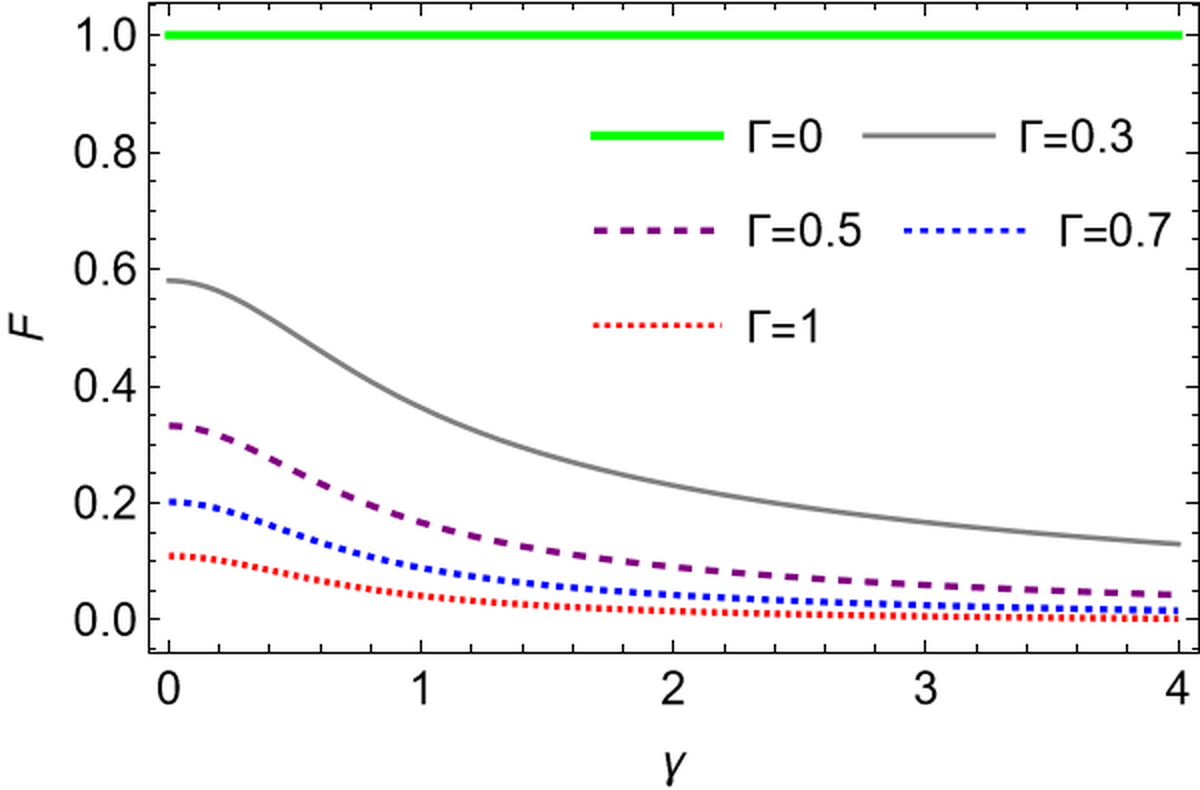}\caption{\label{fig:11}Fidelity $F$ as a function of the state parameter
$\gamma$. Here, we take weak value as $\langle\sigma_{x}\rangle_{w}=5.761$
and other parameters are the same as Fig. \ref{fig:2}. }
\end{figure}

In this sub-section, we examine the state distance between the initial
PCS and final MD states, specifically $\vert\phi\rangle$ and $\vert\Psi\rangle$
using the fidelity function. The fidelity $F$ is defined as the square
of the absolute value of their scalar product, and it is bounded as
$0\leq F\leq1$, where the lower (higher) boundary represents completely
different (identical) states. We define the fidelity function between
$\vert\phi\rangle$ and $\vert\Psi\rangle$ as: 

\begin{equation}
F=|\langle\phi|\Psi\rangle|^{2}.\label{eq:35}
\end{equation}
After substituting $\vert\phi\rangle$ and $\vert\Psi\rangle$ {(}see
Eqs. (\ref{eq:fock}) and (\ref{eq;postselection}){)} into the
ab ove formula, one can derive an explicit expression for $F$. However,
we only present the numerical analysis of this quantity. As shown
in Fig. \ref{fig:11}, the postselected von Neumann measurement indeed
causes a change in the given state, resulting in a larger state distance
between $\vert\phi\rangle$ and $\vert\Psi\rangle$, which even transforms
it into a distinguishable state with increasing coupling strength
parameter $\Gamma$ and state parameter $\gamma$. Larger anomalous
weak values have a more significant effect than smaller ones in distinctly
transforming between the states.

\subsection{Fidelity of teleportation}

As discussed in Sec. \ref{sec:The-effects entanglement}, the postselected
von Neumann measurement can enhance the entanglement between two modes
of PCS for some parameter regions. We know that the distillation or
purification of entanglement are vital to realize the long-standing
quantum communication. In order to check the effects of our von Neumann-type
measurment on the efficiency of teleportation by using the PCS as
the entangled channel, we use the fidelity to measure the performance
of entanglement. 

Here we consider the protocol of quantum teleportation was proposed
by Vaidman, Braunstein and Kimble (called VBK protocol) \citep{PhysRevA.49.1473,PhysRevLett.80.869,706}.
We take the final MD state $\vert\Psi\rangle_{ab}$ {[}see the Eq.
(\ref{eq;postselection}){]} as the teleportation channel to teleport
a coherent state by using VBK protocol. Next, we briefly describe
the teleportation processes. We assume that as the sender and recipient
the Alice and Bob possess and could control the $a$ mode and $b$
mode of the state $\vert\Psi\rangle_{ab}$, respectively, and the
there have a coherent state $\vert\alpha\rangle_{c}$ in $c$ mode
to be teleported. Thus, initially the composite three modes state
takes the form $\vert\Psi_{1}\rangle=\vert\Psi\rangle_{ab}\vert\alpha\rangle_{c}$.
Firstly, Alice take a joint measurement of the orthogonal quadrature
components of modes $a$ and $c$ characterized by \citep{PhysRevA.62.062304}
\begin{equation}
\vert\beta\rangle_{ac}=\frac{1}{\sqrt{\pi}}\sum_{k=0}^{\infty}D_{c}(\beta)\vert k\rangle_{a}\vert k\rangle_{c},\label{eq:39-1}
\end{equation}
where $D_{c}(\beta)$ represents the displacement operator acting
on mode $c$. After the joint measurement the state $\vert\Psi_{1}\rangle$
changes to $\vert\Psi_{2}\rangle={}_{ac}\langle\beta\vert\Psi_{1}\rangle$.
Then, Alice communicate with Bob with classical information channel
and tell the information of the parameter $\beta$. Finally, Bob takes
a displacement operation over his $b$ mode according to the received
value of $\beta$ from Alice. After all operations are completed,
the resulting state at Bob's side is given as 
\begin{align}
\vert\Psi_{b}\rangle & =\frac{\mathcal{N}_{\delta}\lambda}{2\sqrt{\pi}}\sum_{n,m=0}^{\infty}\frac{\gamma^{n+\delta/2}h_{m}}{\sqrt{n!(n+\delta)!}}\nonumber \\
 & \times\left(t_{+}J_{+,mn}+t_{-}J_{-,mn}\right)D_{b}\left(\beta\right)\vert n\rangle_{b},\label{eq:40}
\end{align}
 where, $h_{m}=\langle m\vert D_{c}\left(-\beta\right)\vert\alpha\rangle_{c}$
, $t_{\pm}=1\pm\langle\sigma_{x}\rangle_{w}$ and $J_{\pm,mn}=\langle m\vert D_{a}\left(\pm\frac{\Gamma}{2}\right)\vert n+\delta\rangle_{a}$. 

\begin{figure}
\includegraphics[width=8cm]{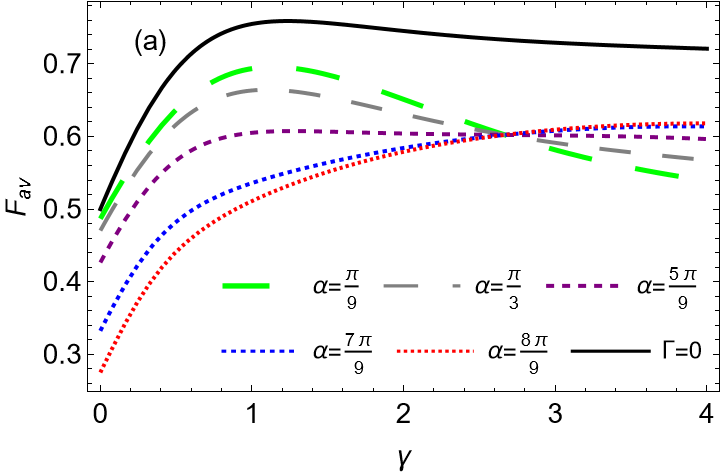}

\includegraphics[width=8cm]{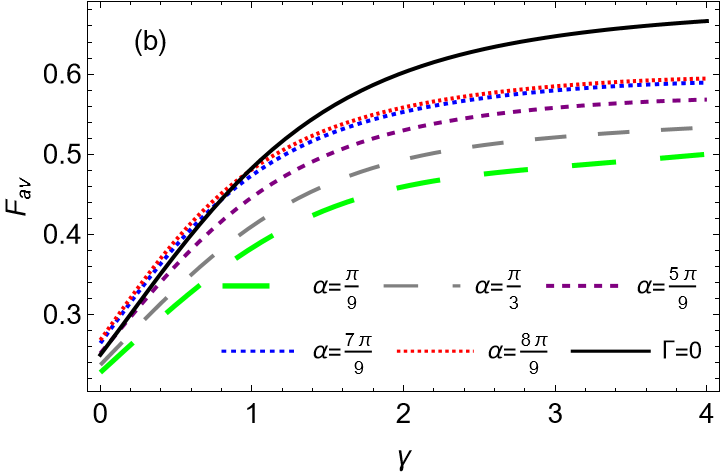}

\includegraphics[width=8cm]{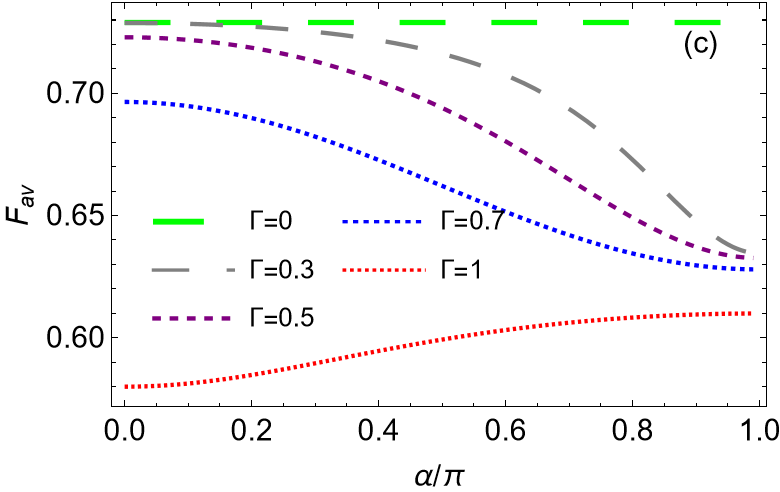}

\caption{\label{fig:12}The fidelity of teleportation of coherent state for
different system parameters. The (a) and (b) present the fidelity
of teleportation as a function of PCS state parameter $\gamma$ for
different values of $\alpha$ . The (c) shows the fidelity of teleportation
as a function of weak value parameter $\alpha$ for different coupling
strength parameter $\Gamma$ while fixed $\gamma=3$ and $\delta=0$.
In (a) $\delta=0$ and (b) $\delta=1$, and $\Gamma=1$ in both cases,
and other parameters are the same as Fig. \ref{fig:2}. }
\end{figure}

To evaluate this teleportation process we can rely on the average
fidelity defined by 
\begin{equation}
F_{av}=\int d^{2}\beta\vert\langle\alpha\vert\Psi_{b}\rangle\vert^{2}.\label{eq:41-1}
\end{equation}
The above integral have exact value and its explicit expression is
listed in Appendix \ref{sec:A1}. The teleportation process is considered
successful when $0.5<F_{av}\le1$ and the higher average fidelity
indicates the better quality of the teleportation process. In Fig.
\ref{fig:12} we plot the average fidelity $F_{av}$ of the teleportation
as a function of $\gamma$ for different weak value parameter $\alpha$.
The black curves in Fig. \ref{fig:12} (a) and (b) corresponding to
the $\delta=0$ and $\delta=1$ cases of initial PCS $\vert\phi\rangle$.
It can observe that the average fidelity $F_{av}$ significantly dependent
on the PND $\delta$. When the PND is set to $\delta=0$, its corresponding
average teleportation fidelity is fully higher than the case of $\delta=1$
for whole parameter regimes. For $\delta=0$ case the maximum value
$0.7589$ of $F_{av}$ achieved at near $\gamma\approx1.22$. 

As showed in Fig. \ref{fig:12} (a), for $\delta=0$ case in small
state parameter $\gamma$ regions the postselected von Neumann measurement
have no positive effect on the teleportation task if the weak value
$\langle\sigma_{x}\rangle_{w}$ takes anomalous values. However, with
increasing the parameter $\gamma$ the average fidelity $F_{av}$
also increases along with large weak values. For $\delta=0$ case,
if $\gamma>0.9$ the average fidelity always larger than the minimum
value ($i.e.,$$0.5$) of successful teleportation and this claim
is valid for all $\Gamma\le1$ regimes {[}see the Fig. \ref{fig:12}
(c){]}. 

In Fig. \ref{fig:12} (b) we presented the average fidelity $F_{av}$
for $\delta=1$ case. It can be seen that in small state parameter
($\gamma$<1) regions the postselection operation could let the $F_{av}$
take larger value than the initial PCS whereas those values can't
guarantee the successful teleportation. However, in $\gamma>1$ regions
of $\delta=1$, same as $\delta=0$ case, after the postselected von
Neumann measurement the average fidelity still can't beyond the initial
PCS. One interesting point in $\delta=1$ case is that with increase
the state parameter $\gamma$ and weak value the $F_{av}$ also increasing,
and can exceed 0.5 for large anomalous weak values and large $\gamma$. 

In a word, compared with the initial PCS state $\vert\phi\rangle$
the postselected von Neumann measurement ($\Gamma\neq0$) can't enhance
the performance of successful teleportation of coherent state if we
take the state $\vert\Psi\rangle_{ab}$ as quantum channel. But, for
PND $\delta=0$ and $\delta=1$ cases, the postselected von Neumann
measurement also could guarantee to keep the successful teleportation
in weak measurement regimes with anomalous weak values. 

\section{\label{sec:Discussion}Discussion}

\begin{figure}
\includegraphics[width=8cm]{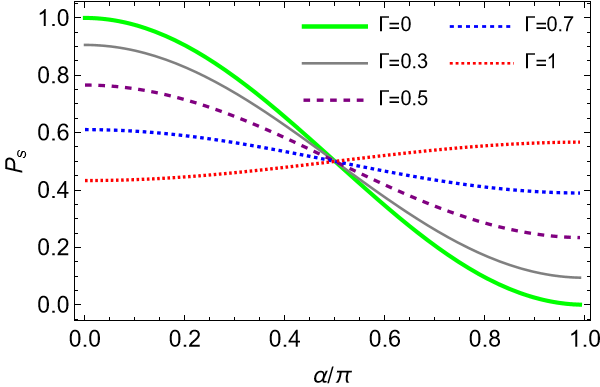}

\caption{\textcolor{blue}{\label{fig:13}The success probability of postselection
of the final MD state $\vert\Psi\rangle$ as a function weak value
parameter $\alpha$ for different coupling strength parameter $\Gamma.$
Here, we take $\delta=0,$ $\gamma=2$ and other parameters are the
same as Fig. \ref{fig:2}.}}

\end{figure}

From the fidelity analysis of our measurement output, the final MD
state $\vert\Psi\rangle$, we can deduce that after the postselected
von Neumann measurement, the initial PCS $\vert\phi\rangle$ undergoes
a dramatic change, potentially transforming into a completely different
state. As previous related studies have shown, the postselected von
Neumann measurement technique provides an alternative approach to
state engineering processes and guarantee the high fidelity \citep{PhysRevA.105.022608}.
In most usual state preparation and state optimization methods the
photon addition and /or photon subtraction are necessary \citep{PhysRevA.73.042310,2007PV,PhysRevA.75.032104,RN170,Chuong2023},
but if one use the postselected von Neumann measurement method to
state preparation processes who can prepare the photon added or photon
subtracted quantum states without adding or subtracting photons \citep{PhysRevA.105.022608}.
In this context, we explore the concrete changes of the final MD state
$\vert\Psi\rangle$ in state preparation processes. Depending on the
coupling strength parameter $\Gamma$ and the state parameter $\gamma$,
we outline its usefulness in the following four scenarios:

(1) Arbitrary values for the state parameter $\gamma$ and the coupling
strength parameter $\Gamma$. In this general case, our final MD state
is the measurement output state $\vert\Psi\rangle$ as defined in
Eq. (\ref{eq;postselection}). For simplicity, the initial PCS $\vert\phi\rangle$
expressed in the coherent state representation as \citep{Agarwal:88}.
\begin{equation}
\vert\phi\rangle=\mathcal{N}_{\delta}e^{^{\vert\gamma\vert}}\int\frac{\mathrm{d}\theta}{2\pi}\left[(\gamma)^{1/2}e^{i\theta}\right]^{-\delta}|(\gamma)^{1/2}e^{i\theta}\rangle_{a}|(\gamma)^{1/2}e^{-i\theta}\rangle_{b}.\label{eq:36}
\end{equation}
Here, $|(\gamma)^{1/2}e^{i\theta}\rangle_{a}=D(\sqrt{\gamma}e^{i\theta})\vert0\rangle_{a}$
and $|(\gamma)^{1/2}e^{-i\theta}\rangle_{b}=D(\sqrt{\gamma}e^{-i\theta})\vert0\rangle_{b}$
denote the coherent states corresponding to the $a$ and $b$ modes
of PCS. In this case, we can rewrite the $\vert\Psi\rangle$ as the
following from 

\begin{align}
\vert\Psi\rangle & =\frac{\lambda}{2}\left[t_{+}|\gamma+\frac{\Gamma_{a}}{2},\delta\rangle+t_{-}|\gamma-\frac{\Gamma_{a}}{2},\delta\rangle\right].\label{eq:37}
\end{align}
Here, the state $|\gamma\pm\frac{\Gamma_{a}}{2},\delta\rangle$ represents
the displaced PCS (the displacement operator $D(\pm\frac{\Gamma}{2})$
acts only on $a$ mode), and its expression is given by: 

\begin{align}
|\gamma\pm\frac{\Gamma_{a}}{2},\delta\rangle & =\frac{\mathcal{N}_{\delta}e^{|\gamma|}}{2\pi}\int\left[(\gamma)^{1/2}e^{i\theta}\right]^{-\delta}e^{\pm\frac{\Gamma}{2}[iIm[((\gamma)^{1/2}e^{-i\theta})^{*}]]}\\
 & \times|(\gamma)^{1/2}e^{i\theta}\pm\frac{\Gamma}{2}\rangle_{a}|(\gamma)^{1/2}e^{-i\theta}\rangle_{b}d\theta.\nonumber 
\end{align}
The final MD state $\vert\Psi\rangle$ is the superposition of $|\gamma+\frac{\Gamma_{a}}{2},\delta\rangle$
and $|\gamma-\frac{\Gamma_{a}}{2},\delta\rangle$. By taking the postselection
probability of measured system (it equals to $\vert\langle\psi_{f}\vert\psi_{i}\rangle\vert^{2}$)
into account, the success probability of postselection of our MD state
is \citep{PhysRevLett.108.080403,PhysRevA.106.022619} 
\begin{align}
P_{s} & =\langle\phi^{\prime}\vert\phi^{\prime}\rangle\nonumber \\
 & =\frac{1}{2}\cos^{2}\frac{\alpha}{2}\left[1+|\langle\sigma_{x}\rangle_{w}|^{2}+(1-|\langle\sigma_{x}\rangle_{w}|^{2})P\right]
\end{align}
where $\vert\phi^{\prime}\rangle=\langle\psi_{f}\vert\Psi_{evol}\rangle$
and the expression of $\vert\Psi_{evol}\rangle$ is given in Eq. (\ref{eq:time evolution}).
For $\delta=0$ case, the dependence of $P_{s}$ on the weak value
$\langle\sigma_{x}\rangle_{w}$ and coupling strength parameter $\Gamma$
for fixed state parameter $\gamma$ is presented in Fig. \ref{fig:13}.
As showed Fig. \ref{fig:13}, the success probability of postselection
of our MD state is not very low especially in very weak coupling and
small weak values regimes. With increase the coupling strength parameter
$\Gamma$ the $P_{s}$ is decreasing for normal weak values of system
observable. Whereas, if $\langle\sigma_{x}\rangle_{w}$ take anomalous
values ($\frac{\pi}{2}$$<\alpha<\pi$), the output state detected
probability $P_{s}$ increased with increasing the coupling strength
$\Gamma$. 

(2) Strong coupling strength parameter ($\Gamma>1$) with small state
parameter $\gamma$ $(\vert\gamma\vert\ll1)$ . Since the state parameter
$\gamma$ is very small, we expand the Eq. (\ref{eq:fock}) up to
the first order of $\gamma$ and ignore higher order terms. In this
extreme case, the Eq. (\ref{eq:37}) changed to: 

\begin{align}
\vert\Psi^{'}\rangle & \thickapprox\frac{\lambda^{'}}{2}\left\{ \left[t_{+}\left|\frac{\Gamma}{2},\delta\right\rangle _{a}+t_{-}\left|-\frac{\Gamma}{2},\delta\right\rangle _{a}\right]\left|0\right\rangle _{b}\right.\nonumber \\
 & +\left.\frac{\gamma}{\sqrt{1+\delta}}\left[t_{+}\left|\frac{\Gamma}{2},1+\delta\right\rangle _{a}+t_{-}\left|-\frac{\Gamma}{2},1+\delta\right\rangle _{a}\right]\left|1\right\rangle _{b}\right\} ,\label{eq:39}
\end{align}
where $\lambda^{'}=\frac{\sqrt{2}\gamma^{\delta/2}}{\sqrt{\delta!}}\left[1+|\langle\sigma_{x}\rangle_{w}|^{2}+(1-|\langle\sigma_{x}\rangle_{w}|^{2})P\right]^{-\frac{1}{2}}$.
The states $\left|\pm\frac{\Gamma}{2},\delta\right\rangle _{a}=D(\pm\frac{\Gamma}{2})\vert\delta\rangle_{a}$
and $\left|\pm\frac{\Gamma}{2},1+\delta\right\rangle _{a}=D(\pm\frac{\Gamma}{2})\vert1+\delta\rangle_{a}$
denote the displaced $\delta$ and $1+\delta$ photon Fock states
of the $a$-mode, respectively. It can be observed that the state
$\vert\Psi^{'}\rangle$ is entangled state of the $a$ and $b$-modes,
and interesting quantum phenomena in the $a$-mode can be explored
by adjusting the related parameters. 

If $\langle\sigma_{x}\rangle_{w}=0$ , i.e. $\alpha=0$, the Eq. (\ref{eq:39})
changed to 
\begin{align}
\vert\Psi_{ch}^{\prime}\rangle\approx & \left\{ \left[\left|\frac{\Gamma}{2},\delta\right\rangle _{a}+\left|-\frac{\Gamma}{2},\delta\right\rangle _{a}\right]\left|0\right\rangle _{b}\right.\nonumber \\
 & +\left.\frac{\gamma}{\sqrt{1+\delta}}\left[\left|\frac{\Gamma}{2},1+\delta\right\rangle _{a}+\left|-\frac{\Gamma}{2},1+\delta\right\rangle _{a}\right]\left|1\right\rangle _{b}\right\} .\label{eq:47}
\end{align}
In this expression the entangled state between two modes can be seen
very clearly. If the $b$-mode is in the vacuum state $\left|0\right\rangle _{b}$,
and the $a$-mode is in superposition of displaced Fock state, i.e.,
$\left|\frac{\Gamma}{2},\delta\right\rangle _{a}+\left|-\frac{\Gamma}{2},\delta\right\rangle _{a}$.
In this case, an interesting point is that if the PND between the
two modes is zero ($\delta=0$), the $a$ mode is prepared in even
Schrödinger cat state $\left|\frac{\Gamma}{2}\right\rangle +\left|-\frac{\Gamma}{2}\right\rangle .$
If we take measurement on mode $b$ and detect a single photon, the
$a$ mode prepared in the superposition of displaced number states:
\begin{equation}
\vert\phi\rangle_{a}\thicksim\langle1\vert_{b}\left(\vert\Psi_{ch}^{\prime}\rangle\right)\thicksim\left[D\left(\frac{\Gamma}{2}\right)+D\left(-\frac{\Gamma}{2}\right)\right]\left|1+\delta\right\rangle _{a},\label{eq:48-1}
\end{equation}

Furthermore, if $\langle\sigma_{x}\rangle_{w}=1$, the above state
became another entangled state as 
\begin{equation}
\vert\psi_{ch}\rangle=\vert\frac{\Gamma}{2},\delta\rangle_{a}\vert0\rangle_{b}+\frac{\gamma}{\sqrt{1+\delta}}\vert\frac{\Gamma}{2},1+\delta\rangle_{a}\vert1\rangle_{b}.\label{eq:48}
\end{equation}
It can be seen that if the $b$-mode in the vacuum state $\vert0\rangle_{b}$
then $a$ mode stated in displaced PND state $D(\frac{\Gamma}{2})\vert\delta\rangle_{a}$
. Upon detecting a single photon at mode $b$, mode $a$ collapses
to displaced $1+\delta$ photon state: 
\begin{equation}
\vert\phi\rangle_{a}\thicksim\langle1\vert_{b}\left(\vert\psi_{ch}\rangle\right)\thicksim D\left(\frac{\Gamma}{2}\right)\vert1+\delta\rangle_{a}.\label{eq:49}
\end{equation}

(3) Weak coupling strength parameter ($\Gamma\ll1$) with state parameter
$\gamma$ ($\vert\gamma\vert>1$). In this case, since the coupling
strength $\Gamma$ is small, we can perform a first-order Taylor expansion
of the associated operators; The displacement operator is approximately
$D(\frac{\Gamma}{2})\approx\mathbb{I}+\frac{\Gamma}{2}\left(a^{\dagger}-a\right)$.
Then, the final MD state $\vert\Psi\rangle$ (Eq. (\ref{eq;postselection}))
changes to (unnormalized): 

\begin{align}
\vert\Psi^{''}\rangle & \thickapprox\left[1+\frac{\Gamma}{2}\left(a^{\dagger}-a\right)Re[\langle\sigma_{x}\rangle_{w}]\right]\vert\phi\rangle,
\end{align}
In this case, the state $\vert\Psi^{''}\rangle$ is in a superposition
of the PCS, single-photon-added PCS, and single photon-subtracted
PCS. Most of the phenomena observed in our current work arise from
this state.

(4) Weak coupling strength parameter ($\Gamma\ll1$) with small state
parameter $\gamma$ ($\vert\gamma\vert<1$). Here, as both the coupling
parameter $\Gamma<1$ and $\gamma$ are very small, we can neglect
$\Gamma\gamma^{1+\delta/2}$ and higher-order terms in our system
state. In this extreme case, the expression for our output state $\vert\Psi\rangle$
further changes to (unnormalized):

\begin{align}
\vert\Psi^{'''}\rangle & \approx\left\{ \left[\left|\delta\right\rangle _{a}+\frac{\Gamma}{2}Re[\langle\sigma_{x}\rangle_{w}]\sqrt{\delta+1}\left|\delta+1\right\rangle _{a}\right.\right.\nonumber \\
 & -\left.\frac{\Gamma}{2}Re[\langle\sigma_{x}\rangle_{w}]\sqrt{\delta}\left|\delta-1\right\rangle _{a}\right]\left|0\right\rangle _{b}\nonumber \\
 & +\left.\frac{\gamma}{\sqrt{1+\delta}}\left|1+\delta\right\rangle _{a}\left|1\right\rangle _{b}\right\} .\label{eq:41}
\end{align}
In this regime, the WVA is dominant, and $\Gamma Re[\langle\sigma_{x}\rangle_{w}]$
can be take big quantity compared to $\gamma$ with low success probability
of postselection, especially if the pre and post-selected system states
are nearly orthogonal. If we assume the PND equal zero ($\delta=0$)
and , the $a$-mode is prepared in a superposition state of $\vert0\rangle_{a}$
and $\vert1\rangle_{a}$, while the $b$-mode remains in the vacuum
state: 
\begin{equation}
\vert\phi^{\prime\prime\prime}\rangle_{a}\thicksim\langle0\vert_{b}\left(\vert\Psi^{\prime\prime\prime}\rangle\right)\thicksim\left|0\right\rangle _{a}+\frac{\Gamma}{2}Re[\langle\sigma_{x}\rangle_{w}]\left|1\right\rangle _{a}.\label{eq:53}
\end{equation}
Furthermore, under sufficiently large anomalous weak value, it is
possible to detect a single photon from the $a$-mode. This feature
could be useful in quantum computation processes, as discussed in
\citep{RevModPhys.79.135}. More interestingly, in this extreme case,
we can also prepare the state $\left|1+\delta\right\rangle _{a}\left|1\right\rangle _{b}$,
but with a lower success probability of postselection. For this extreme
case, the advantages of our scheme are highlighted by the very small
value regimes shown in the numerical results of Figs. \ref{fig:7}
(a) and \ref{fig:7} (b). As we observed, the above prepared states
all strongly depends on PND $\delta$ of PCS, and we can obtain various
displaced photon numbers states and their superposition states just
by changing $\delta$ without photon addition or photon subtraction
processes. 

In recent study \citep{Chuong2023}, the enhancement of non-Gaussianity
and nonclassicality of PCS is investigated by adding $k$ photons
on $a$-mode while subtracting $l$ photons from the b-mode, and showed
that the features including squeezing, quantum correlations and phase
space distributions of initial PCS are improved if $k$ and $l$ both
greater than one (i.e. $k,l\gg1$). Furthermore, in Ref. \citep{Truong2024}
the same task is considered via subtraction photons from two modes
of PCS simultaneously and obtained similar results as in \citep{Chuong2023}.
However, it's important to mention that photon addition and/or subtraction
operation are probabilistic processes, generally having low success
rates even for single photon addition and/or subtraction case. For
example, the non-Gaussian optical states can be generated via heralding
schemes using photon detectors, with typically use beam splitters
to realize the photon subtraction \citep{N2017}. In this case, the
transmissivity parameters must be considered when modeling the photon
subtraction through a beam splitter, which dramatically reduces the
probability of successful state optimization. The accomplishments
of above mentioned multiphoton addition and/or subtraction processes
need more than one beam splitters and higher precision to photon detectors.
Needless to say, such multiphoton addition and/or subtraction operation
based optimization to the PCS have very low success efficiency. 

Compared with previous state optimization methods, our postselected
von Neumann measurement based optimization on PCS is easy to preform
and doesn't need any photon addition and/or subtraction operations.
In our scheme the output state detected probability is not very low
{[}see Fig.\ref{fig:13}{]} and the efficiency also can be controlled
by adjusting the coupling strength parameter and weak values. However,
our state optimization method also has its limitations. With the existence
of postselection process, we can select partial information based
on our wanted outcome and discard the information from the initially
prepared state. On the other hand, the large anomalous weak values
are often accompanied by a low probability of successful postselection.
Due to these limitations, obtaining the desired results via this method
requires a large number of repeated measurements on the prepared systems
as any precision measurements accomplished with postselected weak
measurement \citep{BeamDeflectionPhysRevLett.102.173601,PhysRevLett.117.230801}. 

We also note that our theoretical scheme could be realized on an optical
platform, as the initial system preparation, post-selection of the
measured system, and weak coupling between the measured system and
MD can all be accomplished in optical labs, as in previous WVA experiments
\citep{PhysRevLett.111.033604,PhysRevA.103.032212}. Furthermore,
it may also be possible to implement our scheme in a trapped ion system,
as the von Neumann-type interaction Hamiltonian $H_{int}=g\sigma_{x}\otimes P_{x}$
can be prepared in two-dimensional harmonic oscillators of trapped
ion systems \citep{PhysRevA.100.062111,Nature2020Yi,PhysRevA.108.042601,RN172,PhysRevA.109.032211}. 

Finally, the postselected von Neumann measurement utilizes the technique
of WVA, which enables the PCS to exhibit excellent non-classical properties.
The PCS under postselected von Neumann measurement can show improved
performance in squeezing, quantum statistics, and entanglement by
selecting appropriate parameters. This method may provide a valuable
quantum resource for checking and enhancing the efficiency of associated
quantum tasks based on PCS \citep{PhysRevA.50.2870,Quantumteleportation,Albert_2019}.

\section{\label{sec:conclusion} conclusion}

In this paper, we have proposed a theoretical model to enhance the
non-Gaussianity and nonclassicality of PCS using postselected von
Neumann measurement. We apply a postselected measurement to one mode
(the $a$-mode) of the PCS and investigate related properties, including
squeezing, quantum statistics, and entanglement. The results show
that after the postselected von Neumann measurement, the associated
characteristics of the measurement output state are enhanced in the
WM regime with anomalous weak values. Quadrature squeezing achieves
better-squeezing effects than the PCS over a wide range of state parameters
$\gamma$ under WM. The sum squeezing effect occurs due to WVA for
larger weak values. Numerical results using HZ correlation and EPR
correlation also show that the entanglement between the two modes
of the postselected measurement-enhanced PCS becomes stronger than
the initial state. We confirm these findings by examining the scaled
joint Wigner function for different system parameters. We also analyze
the changes in the initial state after the measurement process in
terms of fidelity and observe that as the coupling strength parameter
$\Gamma$ increases, the initial state $\vert\phi\rangle$ changes
and can even become orthogonal to $\vert\phi\rangle$. From this perspective,
we also discuss the potential applications of our scheme in quantum
state preparation processes and found that various quantum states
with high probability could be generated with some additional measurement
procedures. Interestingly, we also noticed that with our enhanced
PCS the photon addition non-Gaussian states also can be generated
without actual photon addition operations.

Furthermore, as an application example of the PCS under postselected
von Neumann measurement, we also explore the quantum teleportation
protocol by using the postselected measurement-enhanced PCS as quantum
channel. The results showed that in weak measurement regimes the successful
teleportation of a coherent state via our enhanced PCS-based quantum
channel for anomalous weak values could be implemented, but the corresponding
averages fidelity still lower than the initial PCS case. Even though
this drawback, the existence of the postselected von Neumann measurement
could lead the average fidelity be higher than the minimum limit value
for successful quantum teleportation, so that the teleportation of
the associated quantum information with our enhanced PCS-based quantum
channel is still guaranteed. 

We explored the possibility of performing postselected von Neumann
measurements in multimode radiation fields. For simplicity, we only
considered the postselected measurement on one mode of the PCS. It
would be interesting to extend this to postselected measurements on
both modes of two-mode states, such as the two-mode squeezed vacuum
state, to generate interesting states, including entangled superpositions
of two coherent states \citep{doi:10.1126/science.aaf2941}. In future
work, it will also be interesting to study the effects of postselected
von Neumann measurements on other Gaussian and non-Gaussian multipartite
continuous-variable radiation fields \citep{2020quantum,RN11,PRXQuantum.2.030204},
as well as their practical applications and implementations in the
laboratory.
\begin{acknowledgments}
This work was supported by the National Natural Science Foundation
of China (No. 12365005) .
\end{acknowledgments}

\appendix

\section{\label{sec:A1}Related expression}

Here, we derive explicit expressions for the relevant average values
and teleportation fidelity $F_{av}$ associated with our proposed
state $\vert\Psi\rangle$. However, due to their complexity, many
of these expressions are too cumbersome to include in the main text.
Therefore, we have provided them in this appendix for reference.

\begin{widetext}

\begin{align}
\langle a^{\dagger}a\rangle & =\frac{\vert\lambda\vert^{2}}{2}\left[\left(1+|\langle\sigma_{x}\rangle_{w}|^{2}\right)P_{11}+\left(1-|\langle\sigma_{x}\rangle_{w}|^{2}\right)P_{12}\right],\\
P_{11} & =\frac{|\gamma|I_{\delta-1}(2|\gamma|)}{I_{\delta}(2|\gamma|)}+\frac{\Gamma^{2}}{4},\nonumber \\
P_{12} & =\mathcal{N}_{\delta}^{2}\sum\limits_{n=0}^{\infty}\frac{|\gamma|^{2n+\delta}}{n!(n+\delta-1)!}e^{-\frac{\Gamma^{2}}{2}}L_{n+\delta}^{(0)}\left(\Gamma^{2}\right)\nonumber \\
 & -\frac{\Gamma}{2}\mathcal{N}_{\delta}^{2}\sum\limits_{n=0}^{\infty}\frac{|\gamma|^{2n+\delta}}{n!(n+\delta)!}e^{-\frac{\Gamma^{2}}{2}}\Gamma L_{n+\delta}^{(1)}\left(\Gamma^{2}\right)\nonumber \\
 & +\frac{\Gamma}{2}\mathcal{N}_{\delta}^{2}\sum\limits_{n=0}^{\infty}\frac{|\gamma|^{2n+\delta}}{n!(n+\delta)!}e^{-\frac{\Gamma^{2}}{2}}\Gamma L_{n+\delta-1}^{(1)}\left(\Gamma^{2}\right)\nonumber \\
 & +\frac{\Gamma^{2}}{4}\mathcal{N}_{\delta}^{2}\sum\limits_{n=0}^{\infty}\frac{|\gamma|^{2n+\delta}}{n!(n+\delta)!}e^{-\frac{\Gamma^{2}}{2}}L_{n+\delta}^{(0)}\left(\Gamma^{2}\right).\nonumber 
\end{align}

\begin{align}
\langle b^{\dagger}b\rangle & =\frac{\vert\lambda\vert^{2}}{2}\left[\left(1+|\langle\sigma_{x}\rangle_{w}|^{2}\right)P_{21}+\left(1-|\langle\sigma_{x}\rangle_{w}|^{2}\right)P_{22}\right],\\
P_{21} & =\frac{|\gamma|I_{\delta+1}(2|\gamma|)}{I_{\delta}(2|\gamma|)},\nonumber \\
P_{22} & =\mathcal{N}_{\delta}^{2}\sum\limits_{n=0}^{\infty}\frac{|\gamma|^{2n+\delta}}{(n-1)!(n+\delta)!}e^{-\frac{\Gamma^{2}}{2}}L_{n+\delta}^{(0)}\left(\Gamma^{2}\right).\nonumber 
\end{align}

\begin{align}
\langle ab\rangle & =\frac{\vert\lambda\vert^{2}}{2}\left[\left(1+|\langle\sigma_{x}\rangle_{w}|^{2}\right)P_{31}+\left(1-|\langle\sigma_{x}\rangle_{w}|^{2}\right)P_{32}\right],\\
P_{31} & =\gamma,\nonumber \\
P_{32} & =\mathcal{N}_{\delta}\mathcal{N}_{\delta-1}\sum\limits_{n=0}^{\infty}\frac{(\gamma^{*})^{-1}|\gamma|^{2n+\delta}}{(n-1)!(n-1+\delta)!}e^{-\frac{\Gamma^{2}}{2}}L_{n+\delta-1}^{(0)}(\Gamma^{2})\nonumber \\
 & -\frac{\Gamma^{2}}{2}\mathcal{N}_{\delta}\mathcal{N}_{\delta-1}\sum\limits_{n=0}^{\infty}\frac{(\gamma^{*})^{-1}|\gamma|^{2n+\delta}}{(n-1)!(n+\delta)!}e^{-\frac{\Gamma^{2}}{2}}L_{n+\delta-1}^{(1)}(\Gamma^{2}).\nonumber 
\end{align}

\begin{align}
\langle a^{2}b^{2}\rangle & =\frac{\vert\lambda\vert^{2}}{2}\left[\left(1+|\langle\sigma_{x}\rangle_{w}|^{2}\right)P_{41}+\left(1-|\langle\sigma_{x}\rangle_{w}|^{2}\right)P_{42}\right],\\
P_{41} & =\gamma^{2},\nonumber \\
P_{42} & =\mathcal{N}_{\delta}\mathcal{N}_{\delta-2}\sum\limits_{n=0}^{\infty}\frac{(\gamma^{*})^{-2}|\gamma|^{2n+\delta}}{(n-2)!(n-2+\delta)!}e^{-\frac{\Gamma^{2}}{2}}L_{n+\delta-2}^{(0)}\left(\Gamma^{2}\right)\nonumber \\
 & -\Gamma^{2}\mathcal{N}_{\delta}\mathcal{N}_{\delta-2}\sum\limits_{n=0}^{\infty}\frac{(\gamma^{*})^{-2}|\gamma|^{2n+\delta}}{(n-2)!(n-1+\delta)!}e^{-\frac{\Gamma^{2}}{2}}L_{n+\delta-2}^{(1)}\left(\Gamma^{2}\right)\nonumber \\
 & +\frac{\Gamma^{4}}{4}\mathcal{N}_{\delta}\mathcal{N}_{\delta-2}\sum\limits_{n=0}^{\infty}\frac{(\gamma^{*})^{-2}|\gamma|^{2n+\delta}}{(n-2)!(n+\delta)!}e^{-\frac{\Gamma^{2}}{2}}L_{n+\delta-2}^{(2)}\left(\Gamma^{2}\right).\nonumber 
\end{align}

\begin{align}
\langle a^{\dagger}ab^{\dagger}b\rangle & =\frac{\vert\lambda\vert^{2}}{2}\left[\left(1+|\langle\sigma_{x}\rangle_{w}|^{2}\right)P_{51}+\left(1-|\langle\sigma_{x}\rangle_{w}|^{2}\right)P_{52}\right],\\
P_{51} & =|\gamma|^{2}+\frac{\Gamma^{2}}{4}\frac{|\gamma|I_{\delta+1}(2|\gamma|)}{I_{\delta}(2|\gamma|)},\nonumber \\
P_{52} & =\mathcal{N}_{\delta}^{2}\sum\limits_{n=0}^{\infty}\frac{|\gamma|^{2n+\delta}}{(n-1)!(n+\delta-1)!}e^{-\frac{\Gamma^{2}}{2}}L_{n+\delta}^{(0)}\left(\Gamma^{2}\right)\nonumber \\
 & -\frac{\Gamma^{2}}{2}\mathcal{N}_{\delta}^{2}\sum\limits_{n=0}^{\infty}\frac{|\gamma|^{2n+\delta}}{(n-1)!(n+\delta)!}e^{-\frac{\Gamma^{2}}{2}}L_{n+\delta}^{(1)}\left(\Gamma^{2}\right)\nonumber \\
 & +\frac{\Gamma^{2}}{2}\mathcal{N}_{\delta}^{2}\sum\limits_{n=0}^{\infty}\frac{|\gamma|^{2n+\delta}}{(n-1)!(n+\delta)!}e^{-\frac{\Gamma^{2}}{2}}L_{n+\delta-1}^{(1)}\left(\Gamma^{2}\right)\nonumber \\
 & +\frac{\Gamma^{2}}{4}\mathcal{N}_{\delta}^{2}\sum\limits_{n=0}^{\infty}\frac{|\gamma|^{2n+\delta}}{(n-1)!(n+\delta)!}e^{-\frac{\Gamma^{2}}{2}}L_{n+\delta}^{(0)}\left(\Gamma^{2}\right).\nonumber 
\end{align}

\begin{align}
\langle a^{\dagger}b\rangle & =\frac{\vert\lambda\vert^{2}}{2}\left(1-|\langle\sigma_{x}\rangle_{w}|^{2}\right)I_{11},\\
I_{11} & =\Gamma^{2}\mathcal{N}_{\delta}\mathcal{N}_{\delta-1}\sum\limits_{n=0}^{\infty}\frac{(\gamma^{*})^{-1}|\gamma|{}^{2n+\delta}}{(n-1)!(n+\delta)!}e^{-\frac{\Gamma^{2}}{2}}L_{(n+\delta-1)}^{(2)}\left(\Gamma^{2}\right)\nonumber \\
 & -\frac{\Gamma^{2}}{2}\mathcal{N}_{\delta}\mathcal{N}_{\delta-1}\sum\limits_{n=0}^{\infty}\frac{(\gamma^{*})^{-1}|\gamma|{}^{2n+\delta}}{(n-1)!(n+\delta)!}e^{-\frac{\Gamma^{2}}{2}}L_{(n+\delta-1)}^{(1)}\left(\Gamma^{2}\right).\nonumber 
\end{align}

\begin{align}
\langle a^{2}\rangle & =\frac{\vert\lambda\vert^{2}}{2}\left[\left(1+|\langle\sigma_{x}\rangle_{w}|^{2}\right)\frac{\Gamma^{2}}{4}+\left(1-|\langle\sigma_{x}\rangle_{w}|^{2}\right)I_{21}\right],\\
I_{21} & =\Gamma^{2}\mathcal{N}_{\delta}^{2}\sum\limits_{n=0}^{\infty}\frac{|\gamma|^{2n+\delta}}{n!(n+\delta)!}e^{-\frac{\Gamma^{2}}{2}}L_{(n+\delta-2)}^{(2)}\left(\Gamma^{2}\right)\nonumber \\
 & +\Gamma^{2}\mathcal{N}_{\delta}^{2}\sum\limits_{n=0}^{\infty}\frac{|\gamma|^{2n+\delta}}{n!(n+\delta)!}e^{-\frac{\Gamma^{2}}{2}}L_{(n+\delta-1)}^{(1)}\left(\Gamma^{2}\right)\nonumber \\
 & +\frac{\Gamma^{2}}{4}\mathcal{N}_{\delta}^{2}\sum\limits_{n=0}^{\infty}\frac{|\gamma|^{2n+\delta}}{n!(n+\delta)!}e^{-\frac{\Gamma^{2}}{2}}L_{(n+\delta)}^{(0)}\left(\Gamma^{2}\right).\nonumber 
\end{align}

\begin{align}
\langle b^{2}\rangle & =\frac{\vert\lambda\vert^{2}}{2}\left[\left(1-|\langle\sigma_{x}\rangle_{w}|^{2}\right)I_{31}\right],\\
I_{31} & =\mathcal{N}_{\delta}\mathcal{N}_{\delta-2}\sum\limits_{n=0}^{\infty}\frac{(\gamma^{*})^{-2}|\gamma|^{2n+\delta}}{(n-2)!(n+\delta)!}\Gamma^{2}L_{n+\delta-2}^{(2)}\left(\Gamma^{2}\right)e^{-\frac{\Gamma^{2}}{2}}.\nonumber 
\end{align}

\begin{align}
\langle a\rangle & =\frac{\vert\lambda\vert^{2}}{2}\left[\Gamma Re\left[\langle\sigma_{x}\rangle_{w}\right]-2i\Im[\langle\sigma_{x}\rangle_{w}]I_{41}\right],\\
I_{41} & =-\Gamma\mathcal{N}_{\delta}^{2}\sum\limits_{n=0}^{\infty}\frac{|\gamma|^{2n+\delta}}{n!(n+\delta)!}e^{\frac{-\Gamma^{2}}{2}}L_{(n+\delta-1)}^{(1)}\left(\Gamma^{2}\right)\nonumber \\
 & -\frac{\Gamma}{2}\mathcal{N}_{\delta}^{2}\sum\limits_{n=0}^{\infty}\frac{|\gamma|^{2n+\delta}}{n!(n+\delta)!}e^{\frac{-\Gamma^{2}}{2}}L_{(n+\delta)}^{(0)}\left(\Gamma^{2}\right).\nonumber 
\end{align}

\begin{align}
\langle b\rangle & =-i\vert\lambda\vert^{2}Im[\langle\sigma_{x}\rangle_{w}]I_{51},\\
I_{51} & =\Gamma\mathcal{N}_{\delta}\mathcal{N}_{\delta-1}\sum\limits_{n=0}^{\infty}\frac{(\gamma^{*})^{-1}|\gamma|^{2n+\delta}}{(n-1)!(n+\delta)!}e^{-\frac{\Gamma^{2}}{2}}L_{(n+\delta-1)}^{(1)}\left(\Gamma^{2}\right).\nonumber 
\end{align}

\begin{align}
\langle a^{\dagger2}a^{2}\rangle & =\frac{\vert\lambda\vert^{2}}{2}\left[\left(1+|\langle\sigma_{x}\rangle_{w}|^{2}\right)K_{11}+\left(1-|\langle\sigma_{x}\rangle_{w}|^{2}\right)K_{12}\right],\\
K_{11} & =\frac{|\gamma|^{2}I_{\delta-2}(2|\gamma|)}{I_{\delta}(2|\gamma|)}+\Gamma^{2}\frac{|\gamma|I_{\delta-1}(2|\gamma|)}{I_{\delta}(2|\gamma|)}+\frac{\Gamma^{4}}{16},\nonumber \\
K_{12} & =\mathcal{N}_{\delta}^{2}\sum\limits_{n=0}^{\infty}\frac{|\gamma|^{2n+\delta}}{n!(n+\delta-2)!}e^{-\frac{\Gamma^{2}}{2}}L_{n+\delta}^{(0)}\left(\Gamma^{2}\right)\nonumber \\
 & -\Gamma\mathcal{N}_{\delta}^{2}\sum\limits_{n=0}^{\infty}\frac{|\gamma|^{2n+\delta}}{n!(n+\delta-1)!}e^{\frac{-\Gamma^{2}}{2}}\Gamma L_{(n+\delta)}^{(1)}\left(\Gamma^{2}\right)+\Gamma\mathcal{N}_{\delta}^{2}\sum\limits_{n=0}^{\infty}\frac{|\gamma|^{2n+\delta}}{n!(n+\delta)!}(n+\delta-1)e^{\frac{-\Gamma^{2}}{2}}\Gamma L_{(n+\delta-1)}^{(1)}\left(\Gamma^{2}\right)\nonumber \\
 & +\frac{\Gamma^{2}}{4}\mathcal{N}_{\delta}^{2}\sum\limits_{n=0}^{\infty}\frac{|\gamma|^{2n+\delta}}{n!(n+\delta)!}e^{-\frac{\Gamma^{2}}{2}}\Gamma^{2}L_{(n+\delta)}^{(2)}\left(\Gamma^{2}\right)+\frac{\Gamma^{2}}{4}\mathcal{N}_{\delta}^{2}\sum\limits_{n=0}^{\infty}\frac{|\gamma|^{2n+\delta}}{n!(n+\delta)!}e^{\frac{-\Gamma^{2}}{2}}\Gamma^{2}L_{(n+\delta-2)}^{(2)}\left(\Gamma^{2}\right)\nonumber \\
 & +\Gamma^{2}\mathcal{N}_{\delta}^{2}\sum\limits_{n=0}^{\infty}\frac{|\gamma|^{2n+\delta}}{n!(n+\delta-1)!}e^{\frac{-\Gamma^{2}}{2}}L_{(n+\delta)}^{(0)}\left(\Gamma^{2}\right)-\frac{\Gamma^{3}}{4}\mathcal{N}_{\delta}^{2}\sum\limits_{n=0}^{\infty}\frac{|\gamma|^{2n+\delta}}{n!(n+\delta)!}e^{\frac{-\Gamma^{2}}{2}}\Gamma L_{(n+\delta)}^{(1)}\left(\Gamma^{2}\right)\nonumber \\
 & +\frac{\Gamma^{3}}{4}\mathcal{N}_{\delta}^{2}\sum\limits_{n=0}^{\infty}\frac{|\gamma|^{2n+\delta}}{n!(n+\delta)!}e^{\frac{-\Gamma^{2}}{2}}\Gamma L_{(n+\delta-1)}^{(1)}\left(\Gamma^{2}\right)+\frac{\Gamma^{4}}{16}\mathcal{N}_{\delta}^{2}\sum\limits_{n=0}^{\infty}\frac{|\gamma|^{2n+\delta}}{n!(n+\delta)!}e^{\frac{-\Gamma^{2}}{2}}L_{(n+\delta)}^{(0)}\left(\Gamma^{2}\right).\nonumber 
\end{align}

\begin{align}
\langle b^{\dagger2}b^{2}\rangle & =\frac{\vert\lambda\vert^{2}}{2}\left[\left(1+|\langle\sigma_{x}\rangle_{w}|^{2}\right)K_{21}+\left(1-|\langle\sigma_{x}\rangle_{w}|^{2}\right)K_{22}\right],\\
K_{21} & =\frac{|\gamma|^{2}I_{\delta+2}(2|\gamma|)}{I_{\delta}(2|\gamma|)},\nonumber \\
K_{22} & =\mathcal{N}_{\delta}^{2}\sum\limits_{n=0}^{\infty}\frac{|\gamma|^{2n+\delta}}{(n-2)!(n+\delta)!}e^{-\frac{\Gamma^{2}}{2}}L_{(n+\delta)}^{(0)}\left(\Gamma^{2}\right).\nonumber 
\end{align}
 The explicit expression of average fidelity $F_{av}$ corresponding
to the Eq. (\ref{eq:41-1}) is given by 

\begin{align}
F_{av} & =\frac{\mathcal{N}_{\delta}^{2}\lambda^{2}}{8}\sum_{n=0}^{\infty}\sum_{m=0}^{\infty}\sum_{k=0}^{\infty}\sum_{l=0}^{\infty}\frac{\vert\gamma\vert^{n+\delta/2}\vert\gamma\vert^{k+\delta/2}}{\sqrt{n!(n+\delta)!k!(k+\delta)!}}\frac{1}{\sqrt{n!m!l!k!}}\frac{1}{2^{m+k}}\left(m+k\right)!\delta_{m+k,n+l}\nonumber \\
{\color{red}} & \times\exp\left(-\left|\frac{\Gamma}{2}\right|^{2}\right)(-1)^{\max(0,l-(\delta+k))}\sqrt{\frac{\min(l,\delta+k)!\min(m,\delta+n)!}{\max(l,\delta+k)!\max(m,\delta+n)!}}L_{\min(l,k+\delta)}^{|l-(k+\delta)|}\left(\left|\frac{\Gamma}{2}\right|^{2}\right)L_{\min(m,n+\delta)}^{|m-(n+\delta)|}\left(\left|\frac{\Gamma}{2}\right|^{2}\right)\nonumber \\
{\color{red}} & \times\left\{ \vert t_{+}\vert^{2}\left(-\frac{\Gamma}{2}\right)^{|l-(k+\delta)|}\left(\frac{\Gamma}{2}\right)^{|m-(n+\delta)|}+t_{+}t_{-}^{\ast}\left(\frac{\Gamma}{2}\right)^{|l-(k+\delta)|}\left(\frac{\Gamma}{2}\right)^{|m-(n+\delta)|}\right.\nonumber \\
{\color{red}} & +t_{+}^{\ast}t_{-}\left(-\frac{\Gamma}{2}\right)^{|l-(k+\delta)|}\left(-\frac{\Gamma}{2}\right)^{|m-(n+\delta)|}+\vert t_{-}\vert^{2}\left.\left(\frac{\Gamma}{2}\right)^{|l-(k+\delta)|}\left(-\frac{\Gamma}{2}\right)^{|m-(n+\delta)|}\right\} 
\end{align}

\end{widetext}

\bibliographystyle{apsrev4-1}
\bibliography{Refs1}

\end{document}